\documentclass[a4paper,11pt]{article}
\usepackage{jheppub}

\usepackage{amssymb,amsmath}
\usepackage[normalem]{ulem}
\usepackage[utf8]{inputenc}
\usepackage{slashed}
\usepackage{here}
\usepackage{csquotes}
\usepackage{comment}
\usepackage{mathrsfs}
\usepackage{float}
\usepackage{multirow}
\usepackage{longtable}
\usepackage[italicdiff]{physics}

\usepackage{tikz}
\usetikzlibrary{arrows.meta,calc}

\usepackage[hang,small,bf]{caption}
\usepackage[subrefformat=parens]{subcaption}
\newcommand{\im}{\mathrm{i}}

\makeatletter
\newcommand*\rel@kern[1]{\kern#1\dimexpr\macc@kerna}
\newcommand*\widebar[1]{%
  \begingroup
  \def\mathaccent##1##2{%
    \rel@kern{0.8}%
    \overline{\rel@kern{-0.8}\macc@nucleus\rel@kern{0.2}}%
    \rel@kern{-0.2}%
  }%
  \macc@depth\@ne
  \let\math@bgroup\@empty \let\math@egroup\macc@set@skewchar
  \mathsurround\z@ \frozen@everymath{\mathgroup\macc@group\relax}%
  \macc@set@skewchar\relax
  \let\mathaccentV\macc@nested@a
  \macc@nested@a\relax111{#1}%
  \endgroup
}
\makeatother

\numberwithin{equation}{section}
\preprint{
\begin{minipage}{5cm}
\small
\flushright
KYUSHU-HET-363
\end{minipage}}

\title{Lattice chiral non-Abelian gauge symmetry

via bosonization}

\author{Soma Onoda}
\affiliation{
Department of Physics, Kyushu University, 744 Motooka, Nishi-ku, Fukuoka 819-0395, Japan\\
}

\emailAdd{onoda.soma@phys.kyushu-u.ac.jp}

\abstract{
A central issue in lattice formulations of chiral gauge theories is how the anomaly
cancellation mechanism of the continuum theory can be realized at finite lattice
spacing. In the present paper, based on non-Abelian bosonization, we propose a
lattice formulation of the bosonic theory corresponding to a two-dimensional
non-Abelian chiral gauge theory. In the continuum theory, the gauge anomaly of
chiral fermions is represented, in the bosonized description, as anomaly inflow
from a three-dimensional Chern--Simons-type bulk contribution contained in a
gauged Wess--Zumino--Witten model.
Motivated by this structure, we introduce gauge-neutral spectator fermions and
use the resulting bosonized description. We then construct a lattice counterpart
of the gauged Wess--Zumino--Witten model with a three-dimensional bulk
extension under appropriate smoothness conditions. A salient feature of this
lattice formulation is the cancellation of the left and right bulk contributions in
the exponentiated action. This cancellation occurs even before taking the
continuum limit when the anomaly-free condition is satisfied, namely when the
left and right representations have identical quadratic indices. Thus, the present
construction realizes the anomaly-cancellation mechanism at finite lattice
spacing via the bosonized description of two-dimensional anomaly-free chiral
gauge theories. Establishing the desired continuum limit remains an important
open problem.
}

\makeatletter
\gdef\@fpheader{}
\makeatother

\begin{document}

\maketitle


\section{Introduction}

A non-perturbative formulation of chiral gauge theories, including the Standard Model,
has been a long-standing fundamental problem in lattice gauge theory~\cite{Luscher:revisited}.
In the continuum theory, fermions coupled chirally to gauge fields generally give rise to
gauge anomalies. Since gauge anomalies obstruct maintaining gauge symmetry as a
redundancy of the quantum theory, a consistent chiral gauge theory requires its matter
content to satisfy an anomaly-free condition. The central issue in formulating a chiral
gauge theory on the lattice is therefore how the corresponding gauge anomaly is cancelled
non-perturbatively for anomaly-free matter content%
\footnote{
Perturbatively, it is known that, based on the Ginsparg--Wilson relation, the
cancellation of gauge anomalies for anomaly-free representations can be controlled at
finite lattice spacing
\cite{Suzuki:anomaly-cancellation,Luscher:perturbative}.
}.

The Ginsparg--Wilson relation provided an essential step toward resolving this problem
\cite{Ginsparg:1981bj}. It was introduced as a condition to be satisfied by a lattice
Dirac operator, and the overlap Dirac operator was constructed as an explicit solution
to this relation
\cite{Neuberger:1997fp,Neuberger:1998wv}. Based on the Ginsparg--Wilson relation,
an exact chiral symmetry on the lattice was also formulated
\cite{Luscher:1998pqa}, leading to lattice constructions of anomaly-free Abelian
chiral gauge theories and the electroweak theory
\cite{Luscher:1998du,Kikukawa:electro-weak,Kadoh:2007-GSW}. On the other hand,
for more general non-Abelian chiral gauge theories, it remains a difficult problem to
construct the fermion measure non-perturbatively while preserving locality and gauge
invariance, and thereby to realize exact gauge invariance at finite lattice spacing under
the anomaly-free condition%
\footnote{
For the fermion measure problem in lattice chiral gauge theories based on the
Ginsparg--Wilson relation, see also Ref.~\cite{Shang:2013yu}. The analysis in
Ref.~\cite{Shang:2013yu} is carried out for two-dimensional non-Abelian gauge groups on the
zero-field-strength configuration space.
}.

In the present paper, instead of directly addressing this problem for fermionic chiral
gauge theories, we pursue a lattice formulation of two-dimensional chiral gauge theory via
bosonization~(for a review, see Ref.~\cite{Tong:18}). For the case of the $U(1)$ gauge
group, lattice formulations of two-dimensional chiral gauge theories via Abelian bosonization
have recently been developed
\cite{DeMarco:2023-chiral-boson,Berkowitz:2023exact,
Morikawa:2024-yet-another,Thorngren:2026disentangler,
Seifnashri:Exactly-Solvable}. Let us recall some characteristic features of these
constructions. First, since the fundamental variables are bosonic, the
Nielsen--Ninomiya theorem~\cite{Nielsen-Ninomiya:1980rz,Nielsen-Ninomiya:1981hk,
Nielsen-Ninomiya:1981xu} can be avoided, and exact chiral symmetry can be realized on
the lattice without doublers~\cite{Chandrasekharan-nonlinear:2003wy,
Berkowitz:2023exact}. Another important feature is that the gauge anomaly appears
simply as the variation of the bosonic action
\cite{Berkowitz:2023exact,Morikawa:2024-yet-another}. Therefore, the same anomaly
structure as that of the continuum theory can be realized explicitly at finite lattice
spacing. For anomaly-free theories, exact gauge invariance is preserved as a symmetry
of the action.

The purpose of the present paper is to extend this bosonization approach to two-dimensional chiral
gauge theories with non-Abelian gauge symmetry. According to non-Abelian bosonization
in two-dimensional continuum theory, $N_f$ flavors of massless Dirac fermions correspond to the
level~$1$ $U(N_f)$ Wess--Zumino--Witten~(WZW) model
\cite{Witten:nonabelian-bosonization}. When background gauge fields are coupled, the
corresponding bosonic theory is a gauged WZW model~\cite{Hull:1989jk,Kaymakcalan:1983qq, Sen:1988im}. Its gauge anomaly is described
not only by local terms on the two-dimensional boundary but also as anomaly inflow from a three-dimensional bulk
extension
\cite{Callan:1984sa,Witten:2019anomaly-inflow}. Therefore, to realize the
anomaly structure of non-Abelian chiral gauge theory on the lattice via bosonization, it
is necessary to construct a gauged WZW action with a bulk contribution from lattice
variables. It is also necessary to show that its dependence on the bulk extension cancels
under the anomaly-free condition.

In the continuum theory, this bulk contribution is expressed in terms of a three-dimensional
Chern--Simons term. Therefore, in constructing a gauged WZW action on the lattice,
a method for defining a Chern--Simons-type bulk phase from lattice variables provides
an important clue. A natural precedent in this context is the Seiberg-type lattice
Chern--Simons construction
\cite{Seiberg:latticeCS,Luscher:1981-Topology}. The Seiberg-type construction gives an
important method for constructing, from lattice gauge fields, a phase corresponding to
the Chern--Simons phase and reproducing its gauge-transformation property on the
lattice. On the other hand, its explicit construction can involve ambiguities in the
interpolation based on fractional powers of link variables. As a result, an exceptional
configuration can appear even within a gauge orbit. More precisely, configurations
related by gauge transformations need not all admit the fractional powers required for
the interpolation. Although such configurations form a measure-zero subset of the
configuration space, they cannot be ignored if one aims to realize chiral gauge symmetry
exactly at finite lattice spacing.

A related construction of a lattice WZW term based on Ginsparg--Wilson fermions is
also known
\cite{Fujiwara:2003lattice-WZW}. In that construction, the WZW term is defined
through an interpolation of finite gauge transformations. However, there also exist
exceptional configurations for which such an interpolation becomes ill-defined. These
examples suggest that, when formulating Chern--Simons/WZW-type terms on the lattice,
it is essential to specify the region of lattice configurations on which the interpolation is
well-defined in a way compatible with gauge redundancy.

A further technical difficulty arises from the transformation property of the group-valued
field~$g$ appearing in the standard non-Abelian bosonization. In the usual formulation,
this field transforms under both the left and right chiral rotations. Thus, a direct
coupling of this field to background gauge fields on the lattice would involve both
the left and right gauge fields. In our construction, this would require a continuous
interpolation in which the two background gauge fields are controlled simultaneously.
Such an interpolation is not straightforward to incorporate into the explicit lattice
construction employed in this paper.

We therefore add gauge-neutral spectator fermions to the original theory. This allows us to
reorganize the target chiral theory into two sectors described by group-valued fields
$g_1$ and $g_2$. Each of these fields transforms only from one side. This reorganization
leaves the gauge anomaly structure of the original theory unchanged. At the same time,
it enables us to construct the gauge-invariant coupling to the corresponding background
gauge field sector by sector
\footnote{
For related continuum formulations of Weyl-fermion and chiral bosonization, see
Refs.~\cite{Sonnenschein-chiral-bosons:1988ug,Frishman:1987se,Harada:1990tw}.
The decomposition of a Dirac WZW field into chiral components can be accompanied
by an additional gauge symmetry~\cite{Harada:1990tw}.
Constructing a lattice formulation that keeps this additional gauge symmetry exact
appears to be nontrivial.
 In the present lattice construction, we avoid introducing this extra
gauge redundancy and instead use spectator fermions to obtain one-sided sectors.
}
. In the lattice construction, we impose smoothness conditions
on the resulting gauge-invariant combinations to define the required interpolations
consistently.

The main result of this paper is to construct this gauged WZW action on the lattice
at a finite lattice spacing. This action contains a three-dimensional bulk term. We
then show that the left and right bulk contributions responsible for anomaly inflow
cancel each other under the anomaly-free condition
\begin{align}
    \Tr\left(t^a_L t^b_L\right)
    =
    \Tr\left(t^a_R t^b_R\right).
\end{align}
In this sense, our lattice construction realizes the same gauge anomaly structure as
the continuum target chiral theory through the bosonized description obtained by
reorganizing the original theory with spectator fermions.

This paper is organized as follows. In Sec.~\ref{fermion}, we review the anomaly
structure of the continuum theory. We then introduce spectator fermions and describe
the target theory obtained by non-Abelian bosonization. In Sec.~\ref{sec:lattice_wzw},
we introduce the lattice field contents and the smoothness conditions imposed on the
lattice fields. We construct the gauged WZW action on the lattice including the bulk term. 
We also show that the bulk contribution cancels under the anomaly-free condition.
In Sec.~\ref{sec:observables}, we give a prescription for extracting correlation
functions of the original fermion theory from the bosonized theory with spectator
fermions. Section~\ref{sec:conclusion} is devoted to the conclusion and outlook. In
the appendices, we collect technical details on the definition of fractional powers, the
bounds following from the smoothness conditions, and the existence of smooth interpolations.

\section{Gauge anomaly structure in the continuum}

\subsection{Anomaly-inflow mechanism for fermions}
\label{fermion}

In this section, we review the anomaly structure of the chiral symmetry in the
two-dimensional massless Dirac fermion system to be reproduced on the lattice.
Let $M_2$ be a two-dimensional Euclidean spacetime. We consider $N_f$~massless
Dirac fermions $\psi_i$ and $\bar\psi_i$ with $i=1,\dots,N_f$. We introduce the
flavor multiplets
\begin{align}
    \Psi :=
    \begin{pmatrix}
        \psi_1\\
        \psi_2\\
        \vdots\\
        \psi_{N_f}
    \end{pmatrix},
    \qquad
    \bar\Psi :=
    \begin{pmatrix}
        \bar\psi_1 & \bar\psi_2 & \cdots & \bar\psi_{N_f}
    \end{pmatrix}.
\end{align}
The action of the free theory is given by
\begin{align}
S[\Psi,\bar\Psi]
:=
\int_{M_2} \dd^2x\,
\bar\Psi\, \im\slashed\partial\, \Psi .
\label{eq:free_massless_dirac}
\end{align}
We define the left- and right-handed components by
\begin{align}
&\gamma_3:=\im\gamma_1\gamma_2,
\qquad
P_{L/R}:=\frac{1\mp\gamma_3}{2},
\qquad
\psi_{L/R,i}:=P_{L/R}\psi_i,
\qquad
\bar\psi_{L/R,i}:=\bar\psi_iP_{R/L}.
\label{def:projection}
\end{align}
We also define
\begin{align}
    \Psi_{L/R} :=
    \begin{pmatrix}
        \psi_{L/R,1}\\
        \psi_{L/R,2}\\
        \vdots\\
        \psi_{L/R,N_f}
    \end{pmatrix},
    \qquad
    \bar\Psi_{L/R}:=
    \begin{pmatrix}
        \bar\psi_{L/R,1} &
        \bar\psi_{L/R,2} &
        \cdots &
        \bar\psi_{L/R,N_f}
    \end{pmatrix}.
\end{align}
Since the mass term is absent, the action~\eqref{eq:free_massless_dirac} can be
decomposed into the left- and right-handed parts as
\begin{align}
S[\Psi,\bar\Psi]
=
\int_{M_2} \dd^2x\,
\left[
\bar\Psi_L\, \im\slashed\partial\, \Psi_L
+
\bar\Psi_R\, \im\slashed\partial\, \Psi_R
\right].
\label{eq:free_massless_dirac_chiral}
\end{align}
Therefore, the action~\eqref{eq:free_massless_dirac_chiral} is invariant under
\begin{align}
    &\Psi_L\to L\Psi_L,
    \qquad
    \bar\Psi_L\to\bar\Psi_LL^\dagger,
    \qquad
    L\in U(N_f),
    \notag\\
    &\Psi_R\to R\Psi_R,
    \qquad
    \bar\Psi_R\to\bar\Psi_RR^\dagger,
    \qquad
    R\in U(N_f).
\end{align}
Thus, the free theory has the global $U(N_f)_L\times U(N_f)_R$ symmetry.

In the following, we start from $N_f$~Dirac fermions. Hence both the left- and
right-handed Weyl fermions are $V$-valued Grassmann fields, where
$V\simeq\mathbb{C}^{N_f}$ denotes the vector space of flavor components. In this
paper, the left and right representations of the gauge group are realized as
possibly different representations on the same vector space~$V$. Unless
otherwise stated, the trace appearing below denotes the ordinary matrix trace on
$V$, and we denote it by $\Tr$.

More generally, when a representation acts nontrivially only on a subspace of
$V\simeq\mathbb{C}^{N_f}$, we regard the corresponding generators as embedded
into $M_{N_f}(\mathbb C)$. For example, when the gauge group is $SU(2)$ and it
acts only on $(\psi_{L,1},\psi_{L,2})$ in the fundamental representation for
$N_f=3$, the generators can be embedded as matrices on
$V\simeq\mathbb{C}^3$ by
\begin{align}
    t^a_L
    =
    \begin{pmatrix}
        \sigma^a/2 & 0 \\
        0 & 0
    \end{pmatrix},
    \qquad a=1,2,3.
\end{align}

We now consider gauging a subgroup of the global symmetry. In the main text, we
take the gauge group to be $G=SU(N)$. We let it act on the left- and right-handed
fermions through different representations realized on the same $N_f$-dimensional
flavor space. Namely, we do not introduce general independent
$SU(N)_L\times SU(N)_R$ gauge fields. Instead, we introduce the same
Lie-algebra-valued gauge field
\begin{align}
A_\mu=A_\mu^a t^a
\end{align}
and couple it to the left- and right-handed fermions through these representations.
The generators $t^a$ in the fundamental representation of $\mathfrak{su}(N)$ are
assumed to satisfy
\begin{align}
[t^a,t^b]=\im\sum_c f^{abc}t^c .
\end{align}
We denote the Hermitian generators in the left and right representations by $t_L^a,t_R^a\in M_{N_f}(\mathbb C)$, and they satisfy
\begin{align}
[t_{L/R}^a,t_{L/R}^b]
=
\im\sum_c f^{abc}t_{L/R}^c .
\label{Lie algebra generators}
\end{align}
We define the same abstract gauge field evaluated in the left and right
representations by
\begin{align}
A_\mu^L :=\sum_a A_\mu^a\, t_L^a,
\qquad
A_\mu^R :=\sum_a A_\mu^a\, t_R^a .
\label{left right background field}
\end{align}

The action coupled to the background gauge field and the corresponding partition
function are defined by
\begin{align}
S[\Psi,\bar\Psi,A]
&:=
\int_{M_2} \dd^2x\,
\bar\Psi_L\, \im\slashed D_L\, \Psi_L
+
\bar\Psi_R\, \im\slashed D_R\, \Psi_R,
\\
Z_{M_2}[A]
&:=
\int
\mathcal D\bar\Psi_L\,\mathcal D\bar\Psi_R\,
\mathcal D\Psi_L\,\mathcal D\Psi_R\,
e^{-S[\Psi,\bar\Psi,A]} .
\label{fermion gauged partition}
\end{align}
Here we have written the integration measure as
\begin{align}
\mathcal D\bar\Psi_L\,\mathcal D\bar\Psi_R\,
\mathcal D\Psi_L\,\mathcal D\Psi_R
:=
\prod_{i=1}^{N_f}
\mathcal D\bar\psi_{L,i}\,
\mathcal D\bar\psi_{R,i}\,
\mathcal D\psi_{L,i}\,
\mathcal D\psi_{R,i}.
\end{align}
The left and right Dirac operators are
\begin{align}
&\slashed D_L
:=
\slashed\partial
-\im\slashed A^L,
&
\slashed D_R
:=
\slashed\partial
-\im\slashed A^R .
\label{Dirac ops.}
\end{align}
Thus the same abstract gauge field $A_\mu^a$ is coupled to the left- and
right-handed fermions through, in general, different representations.

Let $\alpha^a(x)\in\mathbb{R}$ be local gauge-transformation parameters, and define
\begin{align}
V_L(x)&:=\exp\left(\im\sum_a\alpha^a(x)t_L^a\right),
&
V_R(x)&:=\exp\left(\im\sum_a\alpha^a(x)t_R^a\right).
\end{align}
Then the fermion fields and the background gauge fields transform as
\begin{align}
    &\Psi_L\to V_L\Psi_L,
    \qquad
    \bar\Psi_L\to\bar\Psi_LV_L^{-1},
    \notag\\
    &\Psi_R\to V_R\Psi_R,
    \qquad
    \bar\Psi_R\to\bar\Psi_RV_R^{-1},
    \notag\\
    &A_\mu^L
    \to
    V_L A_\mu^L V_L^{-1}
    -\im(\partial_\mu V_L)V_L^{-1},
    \notag\\
    &A_\mu^R
    \to
    V_R A_\mu^R V_R^{-1}
    -\im(\partial_\mu V_R)V_R^{-1}.
\label{fermion gauge transf}
\end{align}
For infinitesimal transformations, this gives
\begin{align}
\delta_\alpha A^{L}
&=
\dd\alpha^L+\im[\alpha^L,A^L],
&
\alpha^L&:=\sum_a\alpha^a t_L^a,
\\
\delta_\alpha A^{R}
&=
\dd\alpha^R+\im[\alpha^R,A^R],
&
\alpha^R&:=\sum_a\alpha^a t_R^a .
\label{infinitesimal gauge transf}
\end{align}
The classical action is invariant under this background gauge transformation.
In general, however, the fermion integration measure
$\mathcal D\bar\Psi_L\,\mathcal D\bar\Psi_R\,
\mathcal D\Psi_L\,\mathcal D\Psi_R$
is not invariant according to the Fujikawa method~\cite{Fujikawa-method:1979ay}. This non-invariance of the
measure corresponds to the gauge anomaly of the two-dimensional chiral gauge
theory.

In our convention, the left- and right-handed Weyl fermion measures give
contributions with opposite signs under the background gauge transformation. The
infinitesimal gauge variation of the fermion partition function is
\begin{align}
\delta_\alpha \ln Z_{M_2}[A]
&=
-\frac{\im}{4\pi}
\int_{M_2}
\Tr\left(\alpha^L\, \dd A^L\right)
+
\frac{\im}{4\pi}
\int_{M_2}
\Tr\left(\alpha^R\, \dd A^R\right)
\notag\\
&=
-\frac{\im}{4\pi}
\int_{M_2}
\sum_{a,b}
\alpha^a
\left[
\Tr(t_L^a t_L^b)
-
\Tr(t_R^a t_R^b)
\right]
\dd A^b .
\label{fermion anomaly explicit}
\end{align}
Equation~\eqref{fermion anomaly explicit} is the explicit form of the consistent
anomaly~\cite{Bardeen:1984pm} of the two-dimensional chiral fermions. In particular, the anomaly
coefficient is given by
\begin{align}
\Tr(t_L^a t_L^b)
-
\Tr(t_R^a t_R^b).
\end{align}

This anomaly can be cancelled by anomaly inflow~\cite{Callan:1984sa} from a three-dimensional bulk
Chern--Simons term. We introduce a three-dimensional manifold $M_3$ whose
boundary is
\begin{align}
\partial M_3=M_2 .
\end{align}
The Chern--Simons functionals in the left and right representations are defined by
\begin{align}
CS_{M_3}^{L}[A]
&:=
\frac{\im}{4\pi}
\int_{M_3}
\Tr\left(
A^L\wedge\dd A^L-\frac{2\im}{3}(A^L)^3
\right),
\\
CS_{M_3}^{R}[A]
&:=
\frac{\im}{4\pi}
\int_{M_3}
\Tr\left(
A^R\wedge\dd A^R-\frac{2\im}{3}(A^R)^3
\right).
\label{CS}
\end{align}%
\footnote{
On a nontrivial bundle, the Chern--Simons term is defined patchwise, with
overlap contributions associated with transition functions. These patchwise data
transform under changes of local trivialization by the standard descent
relations. Thus, for the left and right sectors obtained from the same
underlying $SU(N)$~gauge field and transition functions, the
representation-dependence argument below, governed by the quadratic index,
applies to the full Chern--Simons data including the overlap terms.
}
Here and below we use the shorthand notation $A^3:=A\wedge A\wedge A$.

We next check that this Chern--Simons term cancels the fermion anomaly in
Eq.~\eqref{fermion anomaly explicit}. With Hermitian generators, our convention for
the infinitesimal gauge transformation is
\begin{align}
\delta_\alpha A
=
\dd\alpha+\im[\alpha,A].
\end{align}
The Chern--Simons 3-form then transforms as
\begin{align}
\delta_\alpha
\Tr\left(
A\wedge \dd A-\frac{2\im}{3}A^3
\right)
=
\dd\,\Tr\left(\alpha\, \dd A\right).
\label{CS variation local}
\end{align}
Thus, by Stokes' theorem, we obtain
\begin{align}
\delta_\alpha CS_{M_3}^{L}[A]
&=
\frac{\im}{4\pi}
\int_{M_2}
\Tr\left(\alpha^L\, \dd A^L\right),
\\
\delta_\alpha CS_{M_3}^{R}[A]
&=
\frac{\im}{4\pi}
\int_{M_2}
\Tr\left(\alpha^R\, \dd A^R\right).
\label{CS variation LR}
\end{align}
Therefore, the gauge variation of the bulk inflow factor
\begin{align}
\exp\left[
CS_{M_3}^{L}[A]-CS_{M_3}^{R}[A]
\right]
\label{inflow action}
\end{align}
is determined by
\begin{align}
\delta_\alpha
\left(
CS_{M_3}^{L}[A]-CS_{M_3}^{R}[A]
\right)
&=
\frac{\im}{4\pi}
\int_{M_2}
\Tr\left(\alpha^L\, \dd A^L\right)
-
\frac{\im}{4\pi}
\int_{M_2}
\Tr\left(\alpha^R\, \dd A^R\right)
\notag\\
&=
+\frac{\im}{4\pi}
\int_{M_2}
\sum_{a,b}
\alpha^a
\left[
\Tr(t_L^a t_L^b)
-
\Tr(t_R^a t_R^b)
\right]
\dd A^b .
\label{CS inflow anomaly explicit}
\end{align}
Adding Eqs.~\eqref{fermion anomaly explicit} and
\eqref{CS inflow anomaly explicit}, we find
\begin{align}
\delta_\alpha \ln Z_{M_2}[A]
+
\delta_\alpha
\left(
CS_{M_3}^{L}[A]-CS_{M_3}^{R}[A]
\right)
=0 .
\label{anomaly inflow cancellation}
\end{align}
Thus, the gauge anomaly of the two-dimensional boundary fermions is precisely
cancelled by the Chern--Simons term in the three-dimensional bulk. The
combination
\begin{align}
Z_{M_2}[A]\,
\exp\left[
CS_{M_3}^{L}[A]-CS_{M_3}^{R}[A]
\right]
\label{gauge inv partition}
\end{align}
is invariant under the background gauge transformation.

From this viewpoint, anomaly cancellation allows us to remove the bulk dependence
while preserving background gauge invariance. To see this condition explicitly, we
examine the representation dependence of the Chern--Simons 3-form. For a
general representation, writing $A=A^a t^a$, we have
\begin{align}
\Tr\left(
A\wedge \dd A-\frac{2\im}{3}A^3
\right)
&=
\sum_{a,b}
\Tr(t^a t^b)\,
A^a\wedge \dd A^b
-\frac{2\im}{3}
\sum_{a,b,c}
\Tr(t^a t^b t^c)\,
A^a\wedge A^b\wedge A^c .
\label{CS calc 1}
\end{align}
For the cubic term, only the antisymmetric part in the Lie-algebra indices
contributes. In particular, antisymmetrizing the first two indices gives
\begin{align}
\Tr(t^a t^b t^c)\,
A^a\wedge A^b\wedge A^c
&=
\frac{1}{2}
\Tr\left([t^a,t^b]t^c\right)
A^a\wedge A^b\wedge A^c
\notag\\
&=
\frac{\im}{2}
\sum_d f^{abd}\Tr(t^d t^c)
A^a\wedge A^b\wedge A^c .
\label{triple trace antisym}
\end{align}
Thus, the representation dependence of the Chern--Simons 3-form appears only
through the quadratic form~$\Tr(t^a t^b)$.
Since the structure constants $f^{abc}$ are determined by the Lie algebra itself
and are common to all representations, the condition
\begin{align}
\Tr(t_L^a t_L^b)=\Tr(t_R^a t_R^b)
\qquad
\text{for all }a,b
\label{anomaly free condition fermion}
\end{align}
implies
\begin{align}
\Tr\left(
A^L\wedge \dd A^L-\frac{2\im}{3}(A^L)^3
\right)
=
\Tr\left(
A^R\wedge \dd A^R-\frac{2\im}{3}(A^R)^3
\right)
\end{align}
for arbitrary background fields $A^a$. Hence,
\begin{align}
CS_{M_3}^{L}[A]-CS_{M_3}^{R}[A]=0,
\qquad
\exp\left[
CS_{M_3}^{L}[A]-CS_{M_3}^{R}[A]
\right]=1 .
\end{align}
In this case,
\begin{align}
Z_{M_2}[A]\,
\exp\left[
CS_{M_3}^{L}[A]-CS_{M_3}^{R}[A]
\right]
=
Z_{M_2}[A],
\end{align}
and the partition function can be defined as a background-gauge-invariant
two-dimensional theory without introducing any bulk dependence.

We have therefore found that the condition~\eqref{anomaly free condition fermion},
namely
\begin{align}
\Tr(t_L^a t_L^b)=\Tr(t_R^a t_R^b)
\qquad
\text{for all }a,b,
\end{align}
is the cancellation condition for the perturbative gauge anomaly in this
two-dimensional chiral theory. This is the non-Abelian counterpart of the condition
\begin{align}
\sum_\alpha q_{L,\alpha}^2
=
\sum_\alpha q_{R,\alpha}^2
\end{align}
in Abelian chiral gauge theory~(e.g. Ref.~\cite{Morikawa:2024-yet-another}).

\subsection{Target theory and non-Abelian bosonization}

To obtain a lattice formulation corresponding to the fermionic chiral gauge
theory~\eqref{fermion gauged partition}, we first pass to an appropriate
bosonized action by using non-Abelian bosonization and then formulate it on the
lattice. For this purpose, we introduce ``spectator fermions'' and consider the
enlarged partition function
\begin{align}
Z_{M_2}^{+\mathrm{sp}}[A]
&:=
\int
\mathcal D\bar\Psi_L\,\mathcal D\bar\Psi_R\,
\mathcal D\Psi_L\,\mathcal D\Psi_R
\int
\mathcal D\bar\Psi_L^{\text{sp}}\,\mathcal D\bar\Psi_R^{\text{sp}}\,
\mathcal D\Psi_L^{\text{sp}}\,\mathcal D\Psi_R^{\text{sp}}
\notag\\
&\qquad\times
\exp\left\{
-\int_{M_2} \dd^2x\,
\left[
\bar\Psi_L\, \im\slashed D_L\, \Psi_L
+
\bar\Psi_R\, \im\slashed D_R\, \Psi_R
+
\bar\Psi_L^{\text{sp}}\, \im\slashed\partial\, \Psi_L^{\text{sp}}
+
\bar\Psi_R^{\text{sp}}\, \im\slashed\partial\, \Psi_R^{\text{sp}}
\right]
\right\}.
\label{with spectator fermion partition function}
\end{align}
This amounts to adding $N_f$ free Dirac fermions which are completely decoupled
from the background gauge field and the original fermions. The introduction of
these spectator degrees of freedom allows us to obtain a bosonized action which is
suitable for the lattice construction below%
\footnote{
Strictly speaking, the fermion partition function depends on the spin structure~$\eta$.
In a bosonized description, this spin-structure dependence can be incorporated by
using a $\mathbb Z_2$~gauge field and the Arf invariant; see, for example,
Refs.~\cite{Thorngren-Bosonization:2018bhj,Karch-Arf:2019lnn}. For simplicity,
the bosonized theory considered in this paper should be regarded not as the partition
function for a fixed spin structure~$\eta$, but as the theory summed over spin
structures, as in Ref.~\cite{Berkowitz:2023exact}.
}.

Indeed, the partition function~\eqref{with spectator fermion partition function}
factorizes as
\begin{align}
Z_{M_2}^{+\mathrm{sp}}[A]
&=
\underbrace{
\int
\mathcal D\bar\Psi_L^{\text{sp}}\,
\mathcal D\bar\Psi_R^{\text{sp}}\,
\mathcal D\Psi_L^{\text{sp}}\,
\mathcal D\Psi_R^{\text{sp}}\,
\exp\left\{
-\int_{M_2}\dd^2x\,
\left[
\bar\Psi_L^{\text{sp}}\,\im\slashed\partial\,\Psi_L^{\text{sp}}
+
\bar\Psi_R^{\text{sp}}\,\im\slashed\partial\,\Psi_R^{\text{sp}}
\right]
\right\}
}_{\text{spectator factor}}
\notag\\
&\qquad\times
\underbrace{
\int
\mathcal D\bar\Psi_L\,\mathcal D\bar\Psi_R\,
\mathcal D\Psi_L\,\mathcal D\Psi_R\,
\exp\left\{
-\int_{M_2}\dd^2x\,
\left[
\bar\Psi_L\,\im\slashed D_L\,\Psi_L
+
\bar\Psi_R\,\im\slashed D_R\,\Psi_R
\right]
\right\}
}_{\text{original partition function }Z_{M_2}[A]} .
\end{align}
Thus, \(Z_{M_2}^{+\mathrm{sp}}[A]\) differs from the original partition
function~\eqref{fermion gauged partition} only by a field-independent overall
factor.
Moreover, if we restrict the class of observables to those which do not contain
spectator fermions, namely
\begin{align}
    \mathcal{O}
    :=
    \{\text{combinations of $\psi_{L,i}$, $\psi_{R,i}$, and $A_\mu^a$}\},
\end{align}
then the spectator-fermion factor cancels between the numerator and the
denominator of expectation values. We therefore obtain
\begin{align}
\langle\mathcal{O}\rangle_{+\mathrm{sp}}
&=
\langle\mathcal{O}\rangle_{\text{original}} .
\label{aux=original}
\end{align}
Consequently, as long as we consider correlation functions of observables without
spectator fermions, the theory with spectator fermions gives the same correlation
functions as the original theory. In the sense that the physical information of a
quantum field theory is encoded in its correlation functions, the spectator fermions
do not change the target theory in the sector considered here.

Next, we factorize the partition function~\eqref{with spectator fermion partition function} as
\begin{align}
Z_{M_2}^{+\mathrm{sp}}[A]
&=
\left[
\int
\mathcal D\bar\Psi_L\,\mathcal D\Psi_L\,
\mathcal D\bar\Psi_R^{\text{sp}}\,
\mathcal D\Psi_R^{\text{sp}}\,
\exp\left\{
-\int_{M_2} \dd^2x\,
\left[
\bar\Psi_L\,\im\slashed D_L\,\Psi_L
+
\bar\Psi_R^{\text{sp}}\,\im\slashed\partial\,\Psi_R^{\text{sp}}
\right]
\right\}
\right]
\notag\\
&\qquad\times
\left[
\int
\mathcal D\bar\Psi_L^{\text{sp}}\,
\mathcal D\Psi_L^{\text{sp}}\,
\mathcal D\bar\Psi_R\,\mathcal D\Psi_R\,
\exp\left\{
-\int_{M_2} \dd^2x\,
\left[
\bar\Psi_L^{\text{sp}}\,\im\slashed\partial\,\Psi_L^{\text{sp}}
+
\bar\Psi_R\,\im\slashed D_R\,\Psi_R
\right]
\right\}
\right].
\label{factorization partition function}
\end{align}
The two factors can be interpreted as the partition functions of Dirac fermions
whose left- and right-handed Weyl components are arranged as
\begin{align}
&\begin{pmatrix}
            \psi_{L,i}\\
            \psi_{R,i}^{\text{sp}}
        \end{pmatrix},
    &
\begin{pmatrix}
    \psi_{L,i}^{\text{sp}}\\
        \psi_{R,i}
        \end{pmatrix}.
\end{align}
In the first factor, only the left-handed component is coupled to the background
field~$A^L$. In the second factor, only the right-handed component is coupled to
the background field~$A^R$.

With this interpretation, we apply non-Abelian bosonization
\cite{Witten:nonabelian-bosonization} to the first and second factors in
Eq.~\eqref{factorization partition function} separately. We first consider the first
factor. Since it has the anomaly of the left-handed Weyl fermion, a
background-gauge-invariant combination must include the inflow factor as
\begin{align}
&\int
\mathcal D\bar\Psi_L\,\mathcal D\Psi_L\,
\mathcal D\bar\Psi_R^{\text{sp}}\,
\mathcal D\Psi_R^{\text{sp}}\,
\exp\left\{
-\int_{M_2} \dd^2x\,
\left[
\bar\Psi_L\, \im\slashed D_L\, \Psi_L
+
\bar\Psi_R^{\text{sp}}\, \im\slashed\partial\, \Psi_R^{\text{sp}}
\right]
\right\}
\exp\left[CS_{M_3}^{L}[A]\right].
\label{left fermion with inflow}
\end{align}
This combination is the special case of Eq.~\eqref{gauge inv partition} with
$A^R=0$ and is invariant under the background gauge transformation.

We now bosonize the first factor. In the continuum theory, the corresponding
bosonic theory is given, up to local counterterms, by the gauged
Wess--Zumino--Witten~(WZW) model at level~$1$, where the gauge field is treated
as a background field~\cite{Witten:nonabelian-bosonization,Kaymakcalan:1983qq,Hull:1989jk}. We denote its kinetic coupling by $\lambda_*$.
In our normalization,
\begin{align}
\frac{1}{4\lambda_*^2}
\int_{M_2} d^2x\,
\Tr\left[
(D_\mu g)^\dagger D_\mu g
\right],
\qquad
\frac{1}{4\lambda_*^2}
=
\frac{1}{8\pi}.
\end{align}

The bosonized action corresponding to the first factor~\eqref{left fermion with inflow}
is
\begin{align}
&\int_{M_2} \dd^2x\,
\left[
\bar\Psi_L\, \im\slashed D_L\, \Psi_L
+
\bar\Psi_R^{\text{sp}}\, \im\slashed\partial\, \Psi_R^{\text{sp}}
\right]
-CS_{M_3}^{L}[A]
\notag\\
&\longrightarrow
S[g_1,A^L]
:=
\int_{M_2} \dd^2x\,
\frac{1}{4\lambda_*^2}
\sum_\mu
\Tr\left[
(D_\mu^L g_1)^\dagger D_\mu^L g_1
\right]
+
\frac{\im}{12\pi}
\int_{M_3}
\Tr\left[(g_1^{-1}\dd g_1)^3\right]
\notag\\
&\qquad\qquad\qquad\qquad\qquad
+\frac{\im}{4\pi}
\int_{M_2}
\Tr\left(
A^L g_1\,\dd g_1^{-1}
\right)
-CS_{M_3}^{L}[A].
\label{left gauged WZW}
\end{align}
Here we have introduced the bosonic field $g_1(x)\in U(N_f)$ and the covariant
derivative
\begin{align}
D_\mu^L g_1
:=
\partial_\mu g_1-\im A_\mu^L g_1 .
\label{def DL g1}
\end{align}
Under the left gauge transformation
\begin{align}
g_1(x)&\to V_L(x)g_1(x),
\\
A_\mu^L(x)&\to
V_L(x)A_\mu^L(x)V_L(x)^{-1}
-\im(\partial_\mu V_L(x))V_L(x)^{-1},
\end{align}
we have
\begin{align}
D_\mu^L g_1\to V_LD_\mu^L g_1,
\qquad
(D_\mu^L g_1)^\dagger D_\mu^L g_1
\to
(D_\mu^L g_1)^\dagger D_\mu^L g_1 .
\end{align}
Therefore, the kinetic term is invariant under the left gauge transformation.
The remaining combination of the WZW term, the boundary counterterm, and the
Chern--Simons term is defined to reproduce the anomaly inflow of the first
factor~\eqref{left fermion with inflow}%
\footnote{
The field $g_1(x)$ is neutral under the right gauge transformation.
}.

Similarly, we apply non-Abelian bosonization to the second factor in
Eq.~\eqref{factorization partition function}. Since the right-handed Weyl fermion
is coupled to the background field in the second factor, the corresponding inflow
factor is $\exp[-CS_{M_3}^{R}[A]]$. Therefore, the corresponding bosonized action is
\begin{align}
&\int_{M_2} \dd^2x\,
\left[
\bar\Psi_L^{\text{sp}}\, \im\slashed\partial\, \Psi_L^{\text{sp}}
+
\bar\Psi_R\, \im\slashed D_R\, \Psi_R
\right]
+CS_{M_3}^{R}[A]
\notag\\
&\qquad\longrightarrow\quad
S[g_2,A^R]
\notag\\
&:=
\int_{M_2} \dd^2x\,
\frac{1}{4\lambda_*^2}
\sum_\mu
\Tr\left[
(D_\mu^R g_2)^\dagger D_\mu^R g_2
\right]
+
\frac{\im}{12\pi}
\int_{M_3}
\Tr\left[(g_2^{-1}\dd g_2)^3\right]
\notag\\
&\qquad
-\frac{\im}{4\pi}
\int_{M_2}
\Tr\left(
A^R g_2^{-1}\dd g_2
\right)
+CS_{M_3}^{R}[A].
\label{right gauged WZW}
\end{align}
Here we have introduced the bosonic field $g_2(x)\in U(N_f)$ and the covariant
derivative
\begin{align}
D_\mu^R g_2
:=
\partial_\mu g_2+\im g_2 A_\mu^R .
\label{def DR g2}
\end{align}
Under the right gauge transformation
\begin{align}
g_2(x)&\to g_2(x)V_R(x)^{-1},
\\
A_\mu^R(x)&\to
V_R(x)A_\mu^R(x)V_R(x)^{-1}
-\im(\partial_\mu V_R(x))V_R(x)^{-1},
\end{align}
we have
\begin{align}
D_\mu^R g_2\to (D_\mu^R g_2)V_R^{-1},
\qquad
(D_\mu^R g_2)^\dagger D_\mu^R g_2
\to
V_R
\left[
(D_\mu^R g_2)^\dagger D_\mu^R g_2
\right]
V_R^{-1}.
\end{align}
Therefore, the kinetic term is invariant under the right gauge transformation after
taking the trace. The remaining combination of the WZW term, the boundary
counterterm, and the Chern--Simons term is defined to reproduce the anomaly
inflow of the second factor%
\footnote{
The field $g_2(x)$ is neutral under the left gauge transformation.
}.

Combining Eqs.~\eqref{left gauged WZW} and~\eqref{right gauged WZW}, the
bosonized counterpart of the gauge-invariant fermionic partition function with the
bulk Chern--Simons factor is
\begin{align}
Z_{M_2}^{+\mathrm{sp}}[A]\,
\exp\left[
CS_{M_3}^{L}[A]-CS_{M_3}^{R}[A]
\right]
\quad\longrightarrow\quad
Z_{M_3}^{\text{bose}}[A],
\end{align}
where
\begin{align}
Z_{M_3}^{\text{bose}}[A]
:=
\int\mathcal{D}g_1\mathcal{D}g_2\,
e^{-S[g_1,A^L]-S[g_2,A^R]},
\qquad
M_2=\partial M_3 .
\label{full bose partition function}
\end{align}
The Chern--Simons part in the action $S[g_1,A^L]+S[g_2,A^R]$ is
\begin{align}
-\,CS_{M_3}^{L}[A]+CS_{M_3}^{R}[A].
\end{align}
Therefore, the bulk-dependent factor appearing in
$e^{-S[g_1,A^L]-S[g_2,A^R]}$ is
\begin{align}
\exp\left[
CS_{M_3}^{L}[A]-CS_{M_3}^{R}[A]\right]
\end{align}
which agrees with the anomaly-inflow factor on the fermion side.

As in the fermionic theory, this dependence on the three-dimensional bulk
manifold $M_3$ disappears if the anomaly-free condition
\begin{align}
\Tr(t_L^a t_L^b)=\Tr(t_R^a t_R^b)
\qquad
\text{for all }a,b
\label{anomaly free condition bose}
\end{align}
is satisfied. Indeed, under the condition~\eqref{anomaly free condition bose}, we have
\begin{align}
CS_{M_3}^{L}[A]-CS_{M_3}^{R}[A]=0.
\end{align}
Thus, the bulk dependence originating from the background gauge field disappears.
In this case, the bosonic theory~\eqref{full bose partition function} can be defined
as a background-gauge-invariant two-dimensional theory%
\footnote{
The dependence on $M_3$ also appears formally through the ordinary WZW terms
\begin{align}
\frac{\im}{12\pi}
\int_{M_3}
\Tr\left[(g_1^{-1}\dd g_1)^3\right]
+
\frac{\im}{12\pi}
\int_{M_3}
\Tr\left[(g_2^{-1}\dd g_2)^3\right].
\end{align}
This is the usual three-dimensional extension of the WZW term. Since the level is
an integer, the partition function is independent of the choice of extension
\cite{Witten:nonabelian-bosonization}.
}.
Namely,
\begin{align}
Z_{M_3}^{\text{bose}}[A]
&=
Z_{M_2}^{\text{gauge inv.}}[A]
:=
\int\mathcal{D}g_1\mathcal{D}g_2\,
e^{-S[g_1,g_2,A]},
\label{anomaly-free action}
\\
S[g_1,g_2,A]
&:=
\int_{M_2} \dd^2x\,
\frac{1}{4\lambda_*^2}
\sum_\mu
\left\{
\Tr\left[
(D_\mu^L g_1)^\dagger D_\mu^L g_1
\right]
+
\Tr\left[
(D_\mu^R g_2)^\dagger D_\mu^R g_2
\right]
\right\}
\notag\\
&\qquad
+
\frac{\im}{12\pi}
\int_{M_3}
\Tr\left[(g_1^{-1}\dd g_1)^3\right]
+
\frac{\im}{12\pi}
\int_{M_3}
\Tr\left[(g_2^{-1}\dd g_2)^3\right]
\notag\\
&\qquad
+\frac{\im}{4\pi}
\int_{M_2}
\left\{
\Tr\left(
A^L g_1\,\dd g_1^{-1}
\right)
-
\Tr\left(
A^R g_2^{-1}\dd g_2
\right)
\right\}.
\label{anomaly-free partition function}
\end{align}
Equation~\eqref{anomaly-free partition function} gives the continuum gauged WZW
theory satisfying the anomaly-free condition. This is the theory which we will
formulate on the lattice.


\section{Gauged Wess--Zumino--Witten model on the lattice}
\label{sec:lattice_wzw}

In this section, we present the central construction of this paper. The continuum
gauged WZW action
\begin{align}
    &S[g_1,A^L]+S[g_2,A^R]\notag\\
    &\qquad=
    \int_{M_2} d^2x\,
\frac{1}{4\lambda_*^2}
\sum_\mu
\left\{
\Tr\left[
(D_\mu^L g_1)^\dagger D_\mu^L g_1
\right]
+
\Tr\left[
(D_\mu^R g_2)^\dagger D_\mu^R g_2
\right]
\right\}
\notag\\
&\qquad+
\frac{\im}{12\pi}
\int_{M_3}
\left\{
\Tr\left[(g_1^{-1}\dd g_1)^3\right]
+
\Tr\left[(g_2^{-1}\dd g_2)^3\right]
\right\}
\notag\\
&\qquad\qquad
+\frac{\im}{4\pi}
\int_{M_2}
\left\{
\Tr\left(
A^L g_1\,\dd g_1^{-1}
\right)
-
\Tr\left(
A^R g_2^{-1}\dd g_2
\right)
\right\}
-CS_{M_3}^{L}[A]
+CS_{M_3}^{R}[A]
\label{completed anomalous action}
\end{align}
contains not only the kinetic terms and the couplings to the background gauge
fields on the two-dimensional boundary~$M_2$, but also the Wess--Zumino terms
and the Chern--Simons terms on the three-dimensional bulk~$M_3$. Therefore, in
constructing its lattice counterpart, it is not sufficient to discretize only the
field variables on the boundary. We also have to construct the bulk extension and
the corresponding topological terms at a finite lattice spacing. Below, we first
define the lattice field contents and our convention for fractional powers. We then
construct lattice counterparts of the Wess--Zumino/Chern--Simons-type bulk terms
from gauge-invariant combinations by using cube-wise interpolations. Finally, we
show that, when the left and right representations satisfy the anomaly-free
condition
\begin{align}
\Tr(t_L^a t_L^b)=\Tr(t_R^a t_R^b)
\qquad
\text{for all }a,b,
\end{align}
the bulk contributions cancel. In this way, the same gauge-anomaly-cancellation
structure as in the continuum theory is reproduced at a finite lattice spacing.

\subsection{Lattice field contents}

We approximate the two-dimensional manifold~$M_2$ and the three-dimensional
manifold~$M_3$ introduced above by square lattices. In the following, we set the
lattice spacing to unity.

We first take the three-dimensional manifold~$M_3$ and its boundary
two-dimensional manifold~$M_2$ as
\begin{align}
    M_3
 &:=
 \left\{
 \xi=(\xi_1,\xi_2,\xi_3)\in\mathbb{R}^3
 \;\middle|\;
 0\leq \xi_1<L_1,\;
 0\leq \xi_2<L_2,\;
 0\leq \xi_3<\infty
 \right\},
 \\
 M_2
 &:=
 \left\{
 \xi=(\xi_1,\xi_2,\xi_3)\in\mathbb{R}^3
 \;\middle|\;
 0\leq \xi_1<L_1,\;
 0\leq \xi_2<L_2,\;
 \xi_3=0
 \right\}.
\end{align}
The directions~$1$ and~$2$ are periodically identified. Hence $M_2$ is regarded
as a two-dimensional torus~$T^2$. The direction~$3$ is treated as a semi-infinite
direction extending from the boundary into the bulk.

Next, we introduce square-lattice discretizations of $M_3$ and $M_2$. We denote
the three-dimensional lattice by~$\Lambda_3$ and the two-dimensional boundary
lattice by~$\Lambda_2=\partial\Lambda_3$. They are defined by
\begin{align}
 \Lambda_3
 &:=
 \left\{
 n=(n_1,n_2,n_3)\in\mathbb{Z}^3
 \;\middle|\;
 0\leq n_1<L_1,\;
 0\leq n_2<L_2,\;
 0\leq n_3<\infty
 \right\},
 \\
 \Lambda_2
 &:=
 \left\{
 n=(n_1,n_2,n_3)\in\mathbb{Z}^3
 \;\middle|\;
 0\leq n_1<L_1,\;
 0\leq n_2<L_2,\;
 n_3=0
 \right\}.
\end{align}
We impose periodic boundary conditions in the directions~$1$ and~$2$. For
simplicity, we take the direction~$3$ to be semi-infinite%
\footnote{
If the number of lattice sites in the direction~$3$ is taken to be finite, only the
definition of the boundary $M_2$ is modified. The construction described below is unchanged. We define $\Lambda_3$ as above so that $\Lambda_2$ is a two-dimensional square lattice with periodic boundary conditions.
}.
In the following, a lattice site is denoted by~$n\in\Lambda_3$. A boundary lattice
site is a site~$n\in\Lambda_2$, namely a site satisfying $n_3=0$.

For each lattice site~$n\in\Lambda_3$, we define the corresponding
three-dimensional unit cube by
\begin{align}
    c(n)
    :=
    \left\{
    \xi\in M_3
    \;\middle|\;
    0\leq \xi_\mu-n_\mu\leq 1
    \quad
    (\mu=1,2,3)
    \right\} .
\end{align}
We also define an oriented plaquette in the $\mu$-$\nu$ plane by
\begin{align}
    p(n;\mu,\nu)
    :=
    \left\{
    \xi\in c(n)
    \;\middle|\;
    0\leq \xi_\mu-n_\mu\leq 1,\;
    0\leq \xi_\nu-n_\nu\leq 1,\;
    \xi_\rho=n_\rho\quad(\rho\neq\mu,\nu)
    \right\} ,
\end{align}
where $\mu\neq\nu$. For a site~$n\in\Lambda_2$ on the boundary lattice, the
two-dimensional unit square on the boundary is written as
\begin{align}
    p(n;1,2)
    =
    \left\{
    \xi\in M_2
    \;\middle|\;
    0\leq \xi_\mu-n_\mu\leq 1
    \quad
    (\mu=1,2),\;
    \xi_3=0
    \right\}.
\end{align}
In this way, $M_3$ is discretized into three-dimensional cubes, and its
boundary~$M_2$ is discretized into two-dimensional plaquettes.

In the continuum theory, the bosonic fields~$g_1(x)$ and~$g_2(x)$ are degrees of
freedom on the boundary manifold~$M_2=\partial M_3$. On the lattice, we first
introduce boundary site variables
\begin{align}
    g_1(n)\in U(N_f),
    \qquad
    g_2(n)\in U(N_f),
    \qquad
    n\in\Lambda_2 .
\end{align}
In order to define the Wess--Zumino-type terms below, we also extend them to the
bulk lattice~$\Lambda_3$. We denote the extended site variables by the same
symbols,
\begin{align}
    g_1(n)\in U(N_f),
    \qquad
    g_2(n)\in U(N_f),
    \qquad
    n\in\Lambda_3 .
\end{align}
We will show that, when the anomaly-free condition is satisfied, the
exponentiated action is independent of the choice of the bulk extension and can be
described only in terms of the boundary degrees of freedom.

We use nearest-neighbor oriented links~$(nm)$. The oppositely oriented link is
denoted by~$(mn)$. When $m=n+\hat\mu$ with $\mu=1,2,3$, the link~$(nm)$ is
regarded as a positively oriented link.

To introduce the background gauge field, we put an $SU(N)$ variable in the
fundamental representation on each oriented link~$(nm)$:
\begin{align}
    U_{nm}\in SU(N).
\end{align}
We define its images in the left and right representations by
\begin{align}
    U^L_{nm}:=R_L(U_{nm}),
    \qquad
    U^R_{nm}:=R_R(U_{nm}) .
\end{align}
Here $R_{L/R}$ denote the unitary representations of $SU(N)$ associated with
the left and right fermion representations. Given an element $U\in SU(N)$
represented by its fundamental matrix, $R_{L/R}(U)$ denotes the corresponding
unitary matrix acting on the left or right representation space. In particular,
\begin{align}
    R_{L/R}(UV)=R_{L/R}(U)R_{L/R}(V),
    \qquad
    R_{L/R}(U^{\dagger})=R_{L/R}(U)^{\dagger}=R_{L/R}(U)^{-1}.
\end{align}
In terms of the generators of $\mathfrak{su}(N)$, we may write
\begin{align}
    U_{nm}
    &=
    \exp\left(\im\sum_a A_{nm}^a t^a\right),
    \\
    U^L_{nm}
    &=
    \exp\left(\im\sum_a A_{nm}^a t_L^a\right),
    \\
    U^R_{nm}
    &=
    \exp\left(\im\sum_a A_{nm}^a t_R^a\right).
\end{align}
Here $t^a$ are the generators in the fundamental representation, while $t_L^a$
and $t_R^a$ are the generators in the left and right representations. For the
oppositely oriented link, we define
\begin{align}
    U_{mn}=U_{nm}^{-1},
    \qquad
    U^L_{mn}
    =
    (U^L_{nm})^{-1}
    =
    (U^L_{nm})^\dagger,
    \qquad
    U^R_{mn}
    =
    (U^R_{nm})^{-1}
    =
    (U^R_{nm})^\dagger .
\end{align}

We next define local gauge transformations of these lattice variables. At each
site~$n$, we introduce
\begin{align}
    V(n)\in SU(N)
\end{align}
and denote its left and right representations by
\begin{align}
    V_L(n):=R_L(V(n)),
    \qquad
    V_R(n):=R_R(V(n)).
\end{align}
In terms of the generators, these are written as
\begin{align}
    V(n)
    &=
    \exp\left(\im\sum_a \alpha^a(n)t^a\right),
    \\
    V_L(n)
    &=
    \exp\left(\im\sum_a \alpha^a(n)t_L^a\right),
    \\
    V_R(n)
    &=
    \exp\left(\im\sum_a \alpha^a(n)t_R^a\right).
\end{align}
The site variables and the background link variables transform as
\begin{align}
    g_1(n)&\longrightarrow V_L(n)\,g_1(n),
    \\
    g_2(n)&\longrightarrow g_2(n)\,V_R(n)^{-1},
\end{align}
and
\begin{align}
    U_{nm}&\longrightarrow V(n)\,U_{nm}\,V(m)^{-1},
    \\
    U^L_{nm}&\longrightarrow V_L(n)\,U^L_{nm}\,V_L(m)^{-1},
    \\
    U^R_{nm}&\longrightarrow V_R(n)\,U^R_{nm}\,V_R(m)^{-1}.
\end{align}

The following gauge-invariant combinations will be the basic building blocks in
the construction of the action:
\begin{align}
    g_1(n)^\dagger U^L_{nm}g_1(m),
    \qquad
    g_2(n)U^R_{nm}g_2(m)^\dagger .
\end{align}
Indeed, they are invariant under the gauge transformations defined above.

We also define the lattice counterparts of the covariant derivatives
$D_\mu^L g_1$ and $D_\mu^R g_2$ in the continuum theory by
\begin{align}
    D^L g_1(n,\mu)
    &:=
    U^L_{n,n+\hat\mu}\,g_1(n+\hat\mu)-g_1(n),
    \\
    D^R g_2(n,\mu)
    &:=
    g_2(n+\hat\mu)\,U^R_{n+\hat\mu,n}-g_2(n).
\end{align}
They transform as
\begin{align}
    D^L g_1(n,\mu)
    &\longrightarrow
    V_L(n)\,D^L g_1(n,\mu),
    \\
    D^R g_2(n,\mu)
    &\longrightarrow
    D^R g_2(n,\mu)\,V_R(n)^{-1}.
\end{align}

Finally, to define fractional powers in the construction of the lattice
action below, we impose suitable smoothness conditions on the site variables and
the background gauge fields. The detailed definition of the fractional powers used
in this paper is given in Appendix~\ref{app:fractional_power}. For the site variables,
we assume the gauge-covariant smoothness conditions
\begin{align}
    \norm{1-g_1(n)^\dagger U^L_{nm}g_1(m)}<\varepsilon,
    \qquad
    \norm{1-g_2(n)U^R_{nm}g_2(m)^\dagger}<\varepsilon,
    \label{smoothness-condition}
\end{align}
for any nearest-neighbor sites $n,m\in\Lambda_3$. Here $\norm{\cdot}$ denotes
the operator norm.

For the background link variables, we first define the plaquette variable in the
fundamental representation by
\begin{align}
    U_p:=\prod_{(nm)\in\partial p}U_{nm}\in SU(N).
\end{align}
For each plaquette~$p\subset\Lambda_3$, we assume the admissibility condition~\cite{Luscher:1981-Topology}:
\begin{align}
    \norm{1-U_p}<\delta.
    \label{admissibility}
\end{align}
Here the product is taken in the oriented order along the boundary~$\partial p$
of the plaquette~$p$. Sufficient bounds on the parameters $\varepsilon$ and $\delta$ are
given in Appendix~\ref{app:bounds}.

\subsection{Construction of the gauge-invariant lattice action}

We define the lattice counterpart of the continuum action~\eqref{completed anomalous action} by
\begin{align}
    S[g_1,g_2,U]
    &=
    \sum_{n\in\Lambda_2}
    \frac{1}{4\lambda^2}
    \sum_\mu
    \Tr\left[\bigl(D^L g_1(n,\mu)\bigr)^\dagger\bigl(D^L g_1(n,\mu)\bigr)\right]
    \notag\\
    &\qquad+
    \sum_{n\in\Lambda_2}
    \frac{1}{4\lambda^2}
    \sum_\mu
    \Tr\left[\bigl(D^R g_2(n,\mu)\bigr)^\dagger\bigl(D^R g_2(n,\mu)\bigr)\right]
    \notag\\
    &\qquad\qquad+
    \im\sum_{c\in\Lambda_3} k_c^L[g_1,U]
    -
    \im\sum_{c\in\Lambda_3} k_c^R[g_2^\dagger,U] .
    \label{lattice anomalous action}
\end{align}
Here $k_c^L[g_1,U]$ and $k_c^R[g_2^\dagger,U]$ are local
Wess--Zumino/Chern--Simons-type terms associated with each cube~$c$. We first define
$k_c^L[g_1,U]$ for the left sector. The right-sector contribution is obtained in
the same way.

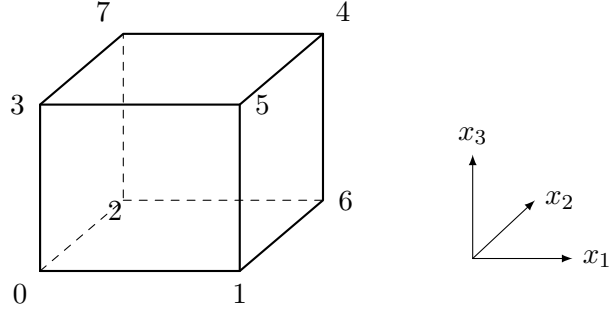
\begin{figure}[t]
\centering
\begin{tikzpicture}[scale=1.1, line cap=round, line join=round]
  \coordinate (v0) at (0,0);
  \coordinate (v1) at (2.4,0);
  \coordinate (v3) at (0,2.0);
  \coordinate (v5) at (2.4,2.0);

  \coordinate (shift) at (1.0,0.85);
  \coordinate (v2) at ($(v0)+(shift)$);
  \coordinate (v6) at ($(v1)+(shift)$);
  \coordinate (v7) at ($(v3)+(shift)$);
  \coordinate (v4) at ($(v5)+(shift)$);

  \draw[thick] (v0)--(v1)--(v6)--(v4)--(v7)--(v3)--(v0);
  \draw[thick] (v3)--(v5)--(v4);
  \draw[thick] (v1)--(v5);

  \draw[dashed] (v0)--(v2)--(v7);
  \draw[dashed] (v2)--(v6);

  \node[below left=1pt]  at (v0) {$0$};
  \node[below=1pt]       at (v1) {$1$};
  \node[left=2pt]        at (v3) {$3$};
  \node[below left=-3.5pt]  at (v2) {$2$};
  \node[above left=1pt]  at (v7) {$7$};
  \node[above right=1pt] at (v4) {$4$};
  \node[right=2pt]       at (v6) {$6$};
  \node[right=2pt]       at (v5) {$5$};

  \coordinate (A) at (5.2,0.15);
  \draw[-latex] (A) -- ++(1.2,0) node[right] {$x_1$};
  \draw[-latex] (A) -- ++(0.75,0.7) node[right] {$x_2$};
  \draw[-latex] (A) -- ++(0,1.25) node[above] {$x_3$};
\end{tikzpicture}
\caption{
Labeling of the vertices of a cube. The coordinate axes $(x_1,x_2,x_3)$ are also
shown. We use
$0=(0,0,0)$, $1=(1,0,0)$, $2=(0,1,0)$, $3=(0,0,1)$,
$4=(1,1,1)$, $5=(1,0,1)$, $6=(1,1,0)$, and $7=(0,1,1)$.
}
\label{fig:cube-label}
\end{figure}

\begin{figure}[t]
\centering
\begin{tikzpicture}[
  scale=1.0,
  line cap=round,
  line join=round,
  facelabel/.style={font=\small},
  vertexlabel/.style={font=\scriptsize, inner sep=1pt}
]
  \def\a{1.6}

  \coordinate (F1) at (0,0);
  \coordinate (F2) at (-\a,0);
  \coordinate (F3) at (0,\a);
  \coordinate (F4) at (0,-\a);
  \coordinate (F5) at (\a,0);
  \coordinate (F6) at (2*\a,0);

  \foreach \F in {F1,F2,F3,F4,F5,F6}{
    \draw[thick] ($(\F)$) rectangle ++(\a,\a);
  }

  \node[facelabel] at ($(F1)+(0.5*\a,0.63*\a)$) {$i=1$};
  \node[facelabel] at ($(F1)+(0.5*\a,0.36*\a)$) {$x_3=0$};

  \node[facelabel] at ($(F2)+(0.5*\a,0.63*\a)$) {$i=2$};
  \node[facelabel] at ($(F2)+(0.5*\a,0.36*\a)$) {$x_2=0$};

  \node[facelabel] at ($(F3)+(0.5*\a,0.63*\a)$) {$i=3$};
  \node[facelabel] at ($(F3)+(0.5*\a,0.36*\a)$) {$x_1=0$};

  \node[facelabel] at ($(F4)+(0.5*\a,0.63*\a)$) {$i=4$};
  \node[facelabel] at ($(F4)+(0.5*\a,0.36*\a)$) {$x_1=1$};

  \node[facelabel] at ($(F5)+(0.5*\a,0.63*\a)$) {$i=5$};
  \node[facelabel] at ($(F5)+(0.5*\a,0.36*\a)$) {$x_2=1$};

  \node[facelabel] at ($(F6)+(0.5*\a,0.63*\a)$) {$i=6$};
  \node[facelabel] at ($(F6)+(0.5*\a,0.36*\a)$) {$x_3=1$};

  \node[vertexlabel, above left]  at ($(F1)+(0,\a)$) {$0$};
  \node[vertexlabel, below left]  at ($(F1)+(0,0)$) {$1$};

  \node[vertexlabel, above left]  at ($(F2)+(0,\a)$) {$3$};
  \node[vertexlabel, below left]  at ($(F2)+(0,0)$) {$5$};

  \node[vertexlabel, above left]  at ($(F3)+(0,\a)$) {$3$};
  \node[vertexlabel, above right] at ($(F3)+(\a,\a)$) {$7$};

  \node[vertexlabel, below left]  at ($(F4)+(0,0)$) {$5$};
  \node[vertexlabel, below right] at ($(F4)+(\a,0)$) {$4$};

  \node[vertexlabel, above left]  at ($(F5)+(0,\a)$) {$2$};
  \node[vertexlabel, below left]  at ($(F5)+(0,0)$) {$6$};

  \node[vertexlabel, above left]  at ($(F6)+(0,\a)$) {$7$};
  \node[vertexlabel, above right] at ($(F6)+(\a,\a)$) {$3$};
  \node[vertexlabel, below left]  at ($(F6)+(0,0)$) {$4$};
  \node[vertexlabel, below right] at ($(F6)+(\a,0)$) {$5$};

\end{tikzpicture}
\caption{
Unfolded net of the cube. Each face is labeled by $i=1,\dots,6$ together with the
corresponding coordinate condition.
}
\label{fig:cube-net-label}
\end{figure}
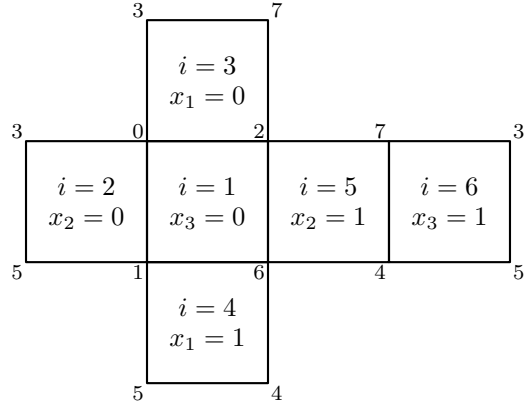

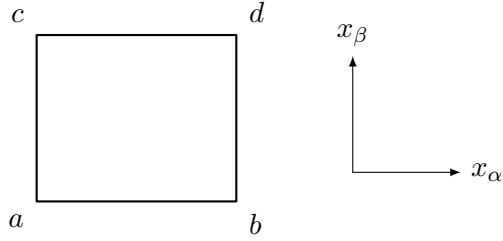
\begin{figure}[t]
\centering
\begin{tikzpicture}[scale=1.1, line cap=round, line join=round]
  \coordinate (pa) at (0,0);
  \coordinate (pb) at (2.4,0);
  \coordinate (pc) at (0,2.0);
  \coordinate (pd) at (2.4,2.0);

  \draw[thick] (pa)--(pb)--(pd)--(pc)--cycle;

  \node[below left=1pt]  at (pa) {$a$};
  \node[below right=1pt] at (pb) {$b$};
  \node[above left=1pt]  at (pc) {$c$};
  \node[above right=1pt] at (pd) {$d$};

  \coordinate (O) at (3.8,0.35);
  \draw[-latex] (O) -- ++(1.3,0) node[right] {$x_\alpha$};
  \draw[-latex] (O) -- ++(0,1.4) node[above] {$x_\beta$};
\end{tikzpicture}
\caption{
Choice of the vertices $(a,b,c,d)$ and the local coordinates
$(x_\alpha,x_\beta)$ on each face. The coordinates are chosen so that
$a=(0,0)$, $b=(1,0)$, $c=(0,1)$, and $d=(1,1)$.
}
\label{fig:face-coord}
\end{figure}

The boundary~$\partial c$ of each cube~$c$ is decomposed into six faces, as shown
in Fig.~\ref{fig:cube-label}. The labels $i=1,\dots,6$ shown in
Fig.~\ref{fig:cube-net-label} denote these faces. For each face, we denote the
corresponding vertices by $(abcd)$ in the oriented order specified in
Fig.~\ref{fig:face-coord}. This order is chosen to be compatible with the
orientation of the boundary of the cube. Explicitly, we take
\begin{align}
    (abcd)
    =
    \begin{cases}
        (0162),&i=1,\\
        (0153),&i=2,\\
        (0273),&i=3,\\
        (1645),&i=4,\\
        (2647),&i=5,\\
        (3547),&i=6.
    \end{cases}
\end{align}

We now define the term $\sum_{c\in\Lambda_3} k_c^L[g_1,U]$ appearing in the
action. For each face $i=1,\dots,6$, we first define
\begin{align}
P_i^L(x_\alpha,x_\beta;g_1)
&:=
\left[g_1(a)^\dagger U_{ac}^L g_1(c)\right]^{x_\beta}
\notag\\
&\times
\Biggl[
\left[g_1(c)^\dagger U_{ca}^L g_1(a)\right]^{x_\beta}
    \,g_1(a)^\dagger
    R_L\left[
        \left(
        U_{ac}U_{cd}U_{db}U_{ba}
        \right)^{x_\beta}
    \right]
    U_{ab}^L g_1(b)
    \left[g_1(b)^\dagger U_{bd}^L g_1(d)\right]^{x_\beta}
\Biggr]^{x_\alpha}.
\label{def:PiL}
\end{align}
Here $a,b,c,d$ denote the vertices of the face~$i$ as shown in
Fig.~\ref{fig:face-coord}. The definition of
$P_i^L(x_\alpha,x_\beta;g_1)$ is gauge invariant by construction.

For $i=4,5,6$, we also define gauge-invariant correction factors by
\begin{align}
R_4^L(x_2,x_3;g_1)
&:=
g_1(0)^\dagger
R_L\Biggl[
\left\{
    \left(
        U_{03} U_{37}U_{72}U_{20}
    \right)^{x_3}
    U_{02}\right.
\notag\\
&\left.\qquad\qquad\quad\times
    \left(
        U_{27}U_{74}U_{46} U_{62}
    \right)^{x_3}
    U_{26} U_{61}
    \left(
        U_{16} U_{64} U_{45}U_{51}
    \right)^{x_3}
    U_{10}\right.
\notag\\
&\left.\qquad\qquad\quad\times
    \left(
        U_{01}U_{15} U_{53} U_{30}
    \right)^{x_3}
\right\}^{x_2}
\left(
    U_{03} U_{35} U_{51} U_{10}
\right)^{x_3}
U_{01}\Biggr]g_1(1),
\\
R_5^L(x_1,x_3;g_1)
&:=
g_1(0)^\dagger
R_L\left[\left(
    U_{03}U_{37} U_{72} U_{20}
\right)^{x_3}
U_{02}\right] g_1(2),
\\
R_6^L(x_1,x_2;g_1)
&:=
g_1(0)^\dagger U_{03}^L g_1(3).
\end{align}
Here the fractional powers of the plaquette variables are defined in the
fundamental representation of $SU(N)$.

We define the interpolation $S^L(x;g_1)$ on the boundary~$\partial c$ by
\begin{align}
    S^L(x;g_1)=P_i^L(x_\alpha,x_\beta;g_1)
    \qquad \text{for } i=1,2,3,
    \label{boundary-SL-123}
\end{align}
and
\begin{align}
    S^L(x;g_1)=R_i^L(x;g_1)\,P_i^L(x_\alpha,x_\beta;g_1)
    \qquad \text{for } i=4,5,6 .
    \label{boundary-SL-456}
\end{align}
The arguments of $R_i^L$ are understood as the appropriate local coordinates on
the corresponding face. Since $P_i^L(x_\alpha,x_\beta;g_1)$ and
$R_i^L(x;g_1)$ are gauge invariant, this definition of
$S^L(x;g_1)$ is also manifestly gauge invariant. Here $x$ is a point on the
corresponding face, and its local coordinates are denoted by
$(x_\alpha,x_\beta)$ or by $(x_1,x_2,x_3)$ depending on the face. We also define
$P^L(x;g_1)$ on each face~$i$ by
\begin{align}
    P^L(x;g_1)=P_i^L(x_\alpha,x_\beta;g_1).
\end{align}

Since $\pi_2(U(N_f))=0$, the boundary field $S^L(x;g_1)$ defined on~$\partial c$
can be extended continuously, and in practice piecewise smoothly, into the interior
of the cube. We denote this extension by the same symbol $S^L(x;g_1)$. The local
contribution $k_c^L[g_1,U]$ associated with the cube~$c$ is then defined by
\begin{align}
    &k_c^L[g_1,U]\notag\\
    &:=
    \frac{1}{12\pi}
    \int_c d^3x
    \sum_{\mu,\nu,\rho}
    \epsilon_{\mu\nu\rho}
    \Tr\left\{
      S^L(x;g_1)\,\partial_\mu (S^L(x;g_1))^{-1}
      \,S^L(x;g_1)\,\partial_\nu (S^L(x;g_1))^{-1}
      \,S^L(x;g_1)\,\partial_\rho (S^L(x;g_1))^{-1}
    \right\}
    \notag\\
    &
    +
    \frac{1}{4\pi}
    \sum_{p\subset\partial c}
    \int_p d^2x\,
    \sum_{\mu,\nu,\rho}
    n_p^\mu\,
    \epsilon_{\mu\nu\rho}\,
    \Tr\left\{
      (P^L(x;g_1))^{-1}\partial_\nu P^L(x;g_1)
      (S^L(x;g_1))^{-1}\partial_\rho S^L(x;g_1)
    \right\}.
    \label{def k_c^L}
\end{align}
Here the sums over $\mu,\nu,\rho$ run over $1,2,3$. In the second term,
$p$ runs over the six plaquettes~(faces) of the cube~$c$, and $n_p^\mu$ denotes the
outward unit normal vector to the plaquette~(face)~$p$. The repeated indices are summed explicitly in
Eq.~\eqref{def k_c^L}. The first term in Eq.~\eqref{def k_c^L} is the local form of
the Wess--Zumino term. Therefore, the ambiguity associated with the extension of
$S^L(x;g_1)$ into the interior of the cube is an addition by an element of
$2\pi\mathbb Z$. Hence this ambiguity does not affect the exponentiated action.
One explicit construction of $S^L(x;g_1)$ inside the cube is given in
Appendix~\ref{S^L-within-cube}. Since $S^L(x;g_1)$ and
$P_i^L(x_\alpha,x_\beta;g_1)$ are gauge invariant, $k_c^L[g_1,U]$ is also gauge
invariant.

The right-sector contribution $k_c^R[g_2^\dagger,U]$ is defined in the same way.
Namely, it is obtained from the left-sector construction by the replacement
\begin{align}
    R_L\to R_R,
    \qquad
    g_1\to g_2^\dagger .
\end{align}
Thus, the Wess--Zumino/Chern--Simons-type term in the lattice action is given by
\begin{align}
    \im\sum_{c\in\Lambda_3} k_c^L[g_1,U]
    -
    \im\sum_{c\in\Lambda_3} k_c^R[g_2^\dagger,U].
\end{align}
This term is manifestly gauge invariant.

\subsection{Anomaly cancellation on the lattice}

In this subsection, we discuss anomaly cancellation in the lattice formulation.
More precisely, we show that, when the anomaly-cancellation condition in the
continuum theory,
\begin{align}
\Tr(t_L^a t_L^b)=\Tr(t_R^a t_R^b)
\qquad
\text{for all }a,b,
\end{align}
is satisfied, the bulk dependence of the lattice action~\eqref{lattice anomalous action}
cancels between the left and right sectors. As a result, the Boltzmann factor can
be written only in terms of degrees of freedom on the boundary~$\Lambda_2=\partial\Lambda_3$.

The part of the lattice action which depends on the degrees of freedom defined in
the three-dimensional bulk~$\Lambda_3$ is the three-dimensional
Wess--Zumino/Chern--Simons-type term. The corresponding part of the Boltzmann factor is
\begin{align}
    \exp\left(
    \im\sum_{c\in\Lambda_3} k_c^L[g_1,U]
    -
    \im\sum_{c\in\Lambda_3} k_c^R[g_2^\dagger,U]
    \right).
    \label{Boltzman}
\end{align}
It is therefore sufficient to show that this factor can be written only in terms
of boundary degrees of freedom under the anomaly-free condition. For this purpose,
we would like to substitute the relation
\begin{align}
    &S^L(x;g_1)=[g_1(0)]^\dagger S^L_I(x;1)g_1^c(x),
    &S^R(x;g_2^\dagger)=g_2(0)S^R_I(x;1)[g_2^c(x)]^\dagger
    \label{gauge tr S}
\end{align}
into Eq.~\eqref{Boltzman}. Since the four objects $S^{L/R}_I(x;1)$ and $g_{1/2}^c(x)$
have not yet been defined, we now define them.

We first define $S^L_I(x;1)$. This object plays the role of a counterpart of the expression obtained from $S^L(x;g_1)$ by formally setting $g_1=1$. However, even if the
link variables $U_{nm}$ satisfy the admissibility condition~\eqref{admissibility},
the condition does not exclude the possibility that a link variable has an
eigenvalue equal to $-1$. In other words, even for an admissible gauge-field
configuration, there can exist link variables for which fractional powers cannot
be defined uniquely. Nevertheless, in the following argument, the object
corresponding to $S^L(x;g_1)|_{g_1=1}$ does not have to be unique. It is enough to
choose one such interpolation for each fixed lattice-field configuration. We use
the notation $S^L_I(x;1)$ to make this interpolation dependence explicit.

We therefore fix a smooth interpolation for each $SU(N)$ matrix~$U$
in the fundamental representation, satisfying the following properties%
\footnote{
For example, suppose that an $SU(N)$-valued matrix~$U$ is diagonalized as
\begin{align}
        U=\gamma\mathrm{diag}(e^{\im\theta_1},\dots,e^{\im\theta_N})\gamma^\dagger,
        \qquad
        \gamma\in SU(N).
\end{align}
Since $U\in SU(N)$, we have
\begin{align}
    \theta_1+\cdots+\theta_N=0 \bmod 2\pi .
\end{align}
If we choose one set of shifts by $2\pi\mathbb Z$,
\begin{align}
      \tilde\theta_i:=\theta_i-2\pi m_i,
      \qquad
      \tilde\theta_1+\cdots+\tilde\theta_N=0,
\end{align}
then an interpolation can be defined by
\begin{align}
     \mathrm{Int}_{x}[U]
     :=
     \gamma\mathrm{diag}(e^{\im x\tilde\theta_1},\dots,e^{\im x\tilde\theta_N})\gamma^\dagger .
\end{align}
}
:
\begin{align}
&\text{For }x\in[0,1],\notag\\
    &\mathrm{Int}_{x}[U]\in SU(N),
    \\
    &\mathrm{Int}_{0}[U]=1,
    \qquad
    \mathrm{Int}_{1}[U]=U.
\end{align}
For link variables, we also impose%
\footnote{
After choosing $\mathrm{Int}_{x}[U_{ab}]$, we define
$\mathrm{Int}_{x}[U_{ba}]$ so that this relation is satisfied.
}
\begin{align}
    \mathrm{Int}_{x}[U_{ba}]
    =
    \left(\mathrm{Int}_{x}[U_{ab}]\right)^{-1}.
\end{align}
The details related to the definition of $\mathrm{Int}_{x}[\cdot]$ are given in
Appendix~\ref{sec:def-Int}. This choice is not unique in general. It is also not required to be smooth as a function of the link variables. Moreover, the explicit form of
$\mathrm{Int}_{x}[U]$ is not gauge invariant. However, for the following argument,
it is sufficient that one such interpolation exists for each fixed configuration of
link variables. The important point is that $\mathrm{Int}_{x}[U]$ is an artificial
interpolation used only in the proof of anomaly cancellation. It is not part of
the definition of the lattice action.

Using this interpolation, we define $S^L_I(x;1)$ on each face of
$\partial c$ as follows:
\begin{align}
P_{i}^L(x_\alpha,x_\beta;1,I)
&:=
R_L\left[
    \mathrm{Int}_{x_\beta}[U_{ac}]
\right.
\notag\\
&\qquad
\times
\left.
\mathrm{Int}_{x_\alpha}
\left[
    \left(
    \mathrm{Int}_{x_\beta}[U_{ac}]
    \right)^{-1}\,
\left(
    U_{ac}U_{cd}U_{db}U_{ba}
\right)^{x_\beta}
U_{ab}\,
\mathrm{Int}_{x_\beta}[U_{bd}]
\right]
\right],
\\
S^L_I(x;1)&:=P_i^L(x_\alpha,x_\beta;1,I)
    \qquad \text{for } i=1,2,3,
\\
S^L_I(x;1)&:=R_i^L(x;1)\,P_i^L(x_\alpha,x_\beta;1,I)
    \qquad \text{for } i=4,5,6.
\end{align}
The arguments of $R_i^L$ are understood as the appropriate local coordinates on
the corresponding face. An explicit extension of this object into the cube is given in
Appendix~\ref{sec:def-Int}. We denote it by the same symbol
$S^L_I(x;1)=S^L_I(x_1,x_2,x_3;1)$.

We now define $g_1^c(x)$ by using $S^L_I(x;1)$:
\begin{align}
    g_1^c(x)&:=\{S^L_I(x;1)\}^{-1}g_1(0)S^L(x;g_1),
    \label{def smooth g_1}
    \\
    S^L(x;g_1)&=[g_1(0)]^\dagger S^L_I(x;1)g_1^c(x).
    \label{gauge tr S^L again}
\end{align}
The important point of the definition~\eqref{def smooth g_1} is the following.
Although $g_1^c(x)$ is defined cube by cube, it gives an interpolation of the
lattice field $g_1(n)$ inside each cube and, furthermore, these interpolations can be glued together to form a $U(N_f)$-valued continuous
and piecewise smooth map on the whole $M_3$. This property is verified in Appendix~\ref{app:continuity-g1c}. It will be crucial
in showing anomaly cancellation.

Now that $S^L_I(x;1)$ and $g_1^c(x)$ have been defined, we substitute
Eq.~\eqref{gauge tr S^L again} into Eq.~\eqref{def k_c^L}, and use the Polyakov--Wiegmann identity. Then, we obtain
\begin{align}
&\sum_{c\in\Lambda_3}k_c^L[g_1,U]\notag\\
    &= \frac{1}{12\pi}\sum_{c\in\Lambda_3}
    \int_c d^3x \sum_{\mu,\nu,\rho}
    \epsilon_{\mu\nu\rho}
    \Tr\{
      (g_1^c(x))^{-1}\,\partial_\mu g_1^c(x)
      \,(g_1^c(x))^{-1}\,\partial_\nu g_1^c(x)
      \,(g_1^c(x))^{-1}\,\partial_\rho g_1^c(x)
    \}
    \notag\\
    &
    +
    \frac{1}{4\pi}\sum_{c\in\Lambda_3}\sum_{p\subset\partial c}
    \int_p d^2x\,
    \sum_{\mu,\nu,\rho}
    n_p^\mu\,
    \epsilon_{\mu\nu\rho}\,
    \Tr\{
      (P^L(x;1,I))^{-1}\partial_\nu P^L(x;1,I)
    \,(g_1^c(x))^{-1}\partial_\rho g_1^c(x)
    \}\notag\\
    &+\frac{1}{12\pi}\sum_{c\in\Lambda_3}
    \int_c d^3x \sum_{\mu,\nu,\rho}
    \epsilon_{\mu\nu\rho}
    \Tr\{
      S_I^L(x;1)\partial_\mu (S_I^L(x;1))^{-1}
      \,S_I^L(x;1)\partial_\nu (S_I^L(x;1))^{-1}
      \,S_I^L(x;1)\,\partial_\rho (S_I^L(x;1))^{-1}
    \}
    \notag\\
    &
    +
    \frac{1}{4\pi}\sum_{c\in\Lambda_3}\sum_{p\subset\partial c}
    \int_p d^2x\,
    \sum_{\mu,\nu,\rho}
    n_p^\mu\,
    \epsilon_{\mu\nu\rho}\,
    \Tr\{
      (P^L(x;1,I))^{-1}\partial_\nu P^L(x;1,I)
      \,(S_I^L(x;1))^{-1}\partial_\rho S_I^L(x;1)
    \}.
   \label{left k_c decomposed}
\end{align}

We first consider the contribution from the first term in
Eq.~\eqref{left k_c decomposed},
\begin{align}
    \frac{1}{12\pi}\sum_{c\in\Lambda_3}
    \int_c d^3x \sum_{\mu,\nu,\rho}
    \epsilon_{\mu\nu\rho}
    \Tr\{
      (g_1^c(x))^{-1}\,\partial_\mu g_1^c(x)
      \,(g_1^c(x))^{-1}\,\partial_\nu g_1^c(x)
      \,(g_1^c(x))^{-1}\,\partial_\rho g_1^c(x)
    \}.
\end{align}
From the discussion in Appendix~\ref{app:continuity-g1c}, $g_1^c(x)$ is defined so that
the interpolations on different cubes are connected together.
Therefore, there exists a continuous map $g_1:M_3\to U(N_f)$ such that
\begin{align}
   g_1(\xi)|_{\xi\in c(n)}=g_1^c(x)|_{x=\xi-n}.
\end{align}
Thus,
\begin{align}
    &\frac{1}{12\pi}
    \sum_{c\in\Lambda_3}\int_c d^3x\sum_{\mu,\nu,\rho}
    \epsilon_{\mu\nu\rho}
    \Tr\{
      (g_1^c(x))^{-1}\,\partial_\mu g_1^c(x)
      \,(g_1^c(x))^{-1}\,\partial_\nu g_1^c(x)
      \,(g_1^c(x))^{-1}\,\partial_\rho g_1^c(x)
    \}\notag\\
    &=
    \frac{1}{12\pi}
    \int_{M_3} d^3\xi\sum_{\mu,\nu,\rho}
    \epsilon_{\mu\nu\rho}
    \Tr\{
      (g_1(\xi))^{-1}\,\partial_\mu g_1(\xi)
      \,(g_1(\xi))^{-1}\,\partial_\nu g_1(\xi)
      \,(g_1(\xi))^{-1}\,\partial_\rho g_1(\xi)
    \}.
    \label{lattice WZ}
\end{align}
This is precisely the Wess--Zumino term in the continuum theory. The standard
quantization argument implies that changing the bulk extension changes this term
only by an element of $2\pi\mathbb Z$. Thus, the exponentiated Boltzmann factor
is determined by the degrees of freedom on the boundary
$M_2=\partial M_3$\footnote{For an explicit derivation of this
standard statement, see, e.g., Refs.~\cite{Gawedzki-gerbe:2002se,Shiozaki-discreate:2024}.}.

Next, we evaluate the contribution from the second term in
Eq.~\eqref{left k_c decomposed},
\begin{align}
    \frac{1}{4\pi}\sum_{c\in\Lambda_3}\sum_{p\subset\partial c}
    \int_p d^2x\,
    \sum_{\mu,\nu,\rho}
    n_p^\mu\,
    \epsilon_{\mu\nu\rho}\,
    \Tr\{
      (P^L(x;1,I))^{-1}\partial_\nu P^L(x;1,I)
    \,(g_1^c(x))^{-1}\partial_\rho g_1^c(x)
    \}.
\end{align}
On each face, $P^L(x;1,I)$ depends only on the link variables on that face and
their interpolations. Hence its value on a shared face is independent of which
adjacent cube is used. The field $g_1^c(x)$ is connected continuously across neighboring
cubes. Therefore, on every shared face in the bulk, the contributions from the two
adjacent cubes cancel each other once the orientations of their boundaries are
taken into account. We obtain
\begin{align}
    &\frac{1}{4\pi}\sum_{c\in\Lambda_3}\sum_{p\subset\partial c}
    \int_p d^2x\,
    \sum_{\mu,\nu,\rho}
    n_p^\mu\,
    \epsilon_{\mu\nu\rho}\,
    \Tr\{
      (P^L(x;1,I))^{-1}\partial_\nu P^L(x;1,I)
    \,(g_1^c(x))^{-1}\partial_\rho g_1^c(x)
    \}\notag\\
    &\qquad=
    \frac{1}{4\pi}\sum_{p\subset\Lambda_2}
    \int_p d^2x\,
    \sum_{\mu,\nu,\rho}
    n_p^\mu\,
    \epsilon_{\mu\nu\rho}\,
    \Tr\{
      (P^L(x;1,I))^{-1}\partial_\nu P^L(x;1,I)
    \,(g_1^c(x))^{-1}\partial_\rho g_1^c(x)
    \}.
    \label{left boundary contribution}
\end{align}
On the right-hand side, $n_p^\mu$ denotes the outward normal of the boundary
plaquette $p\subset\Lambda_2$ regarded as a face of $\Lambda_3$. Thus, this term
depends only on degrees of freedom defined on $\Lambda_2=\partial\Lambda_3$.

The remaining dependence on the bulk~$\Lambda_3$ is the part depending only on
the link variables. It comes from the third and fourth terms in
Eq.~\eqref{left k_c decomposed}:
\begin{align}
    &\frac{1}{12\pi}\sum_{c\in\Lambda_3}
    \int_c d^3x \sum_{\mu,\nu,\rho}
    \epsilon_{\mu\nu\rho}
    \Tr\{
      S_I^L(x;1)\partial_\mu (S_I^L(x;1))^{-1}
      \,S_I^L(x;1)\partial_\nu (S_I^L(x;1))^{-1}
      \,S_I^L(x;1)\,\partial_\rho (S_I^L(x;1))^{-1}
    \}
    \notag\\
    &\qquad
    +
    \frac{1}{4\pi}\sum_{c\in\Lambda_3}\sum_{p\subset\partial c}
    \int_p d^2x\,
    \sum_{\mu,\nu,\rho}
    n_p^\mu\,
    \epsilon_{\mu\nu\rho}\,
    \Tr\{
      (P^L(x;1,I))^{-1}\partial_\nu P^L(x;1,I)
      \,(S_I^L(x;1))^{-1}\partial_\rho S_I^L(x;1)
    \}.
\end{align}

We now show that this term cancels against the right-sector contribution under
the anomaly-cancellation condition. The interpolation $\mathrm{Int}_x$
is chosen by first interpolating the same fundamental-representation link
variables and then mapping the result to the left and right representations by
$R_L$ and $R_R$. Moreover, all fractional powers of pure gauge variables appearing
in the interpolation are defined in the fundamental representation before
being mapped to $R_L$ or $R_R$. Hence $P^L(x;1,I)$ and $S_I^L(x;1)$ are
constructed from the same $SU(N)$ interpolation evaluated in the left
representation. Therefore,
$(S_I^L(x;1))^{-1}\partial_\mu S_I^L(x;1)$ and
$(P^L(x;1,I))^{-1}\partial_\mu P^L(x;1,I)$ can be written as linear combinations
of the generators $t_L^a$%
\footnote{
Let $X(y)\in \mathfrak{su}(N)$. Then
\begin{align}
    e^{-X(y)}\partial_y e^{X(y)}
    =\int_0^1\dd s\,
    e^{-sX(y)}\,\partial_yX(y)\,e^{sX(y)} .
\end{align}
Furthermore, the Campbell--Baker--Hausdorff formula gives
\begin{align}
    e^{-sX(y)}\,\partial_yX(y)\,e^{sX(y)}
    =
    \sum_{n=0}^{\infty}\frac{(-s)^n}{n!}\,
    \mathrm{ad}_{X(y)}^n\bigl(\partial_yX(y)\bigr)
    \in \mathfrak{su}(N).
\end{align}
Indeed, $X(y),\partial_yX(y)\in \mathfrak{su}(N)$, and
$\mathfrak{su}(N)$ is closed under commutators. Hence
$\mathrm{ad}_{X(y)}^n(\partial_yX(y))\in \mathfrak{su}(N)$, and therefore
\begin{align}
    e^{-X(y)}\partial_y e^{X(y)}\in \mathfrak{su}(N).
\end{align}
Thus, if $X(y)=\sum_a X^a(y)t_L^a$, then
$e^{-X(y)}\partial_y e^{X(y)}$ can also be written as a linear combination of
$t_L^a$.
}.
That is, there exist coefficients $\mathcal{A}^a_\mu(x)$ and
$\mathcal{B}^a_\mu(x)$ such that
\begin{align}
    \exists\, \mathcal{A}^a_\mu(x),\mathcal{B}^a_\mu(x)\in\mathbb{C},\qquad
&(S_I^L(x;1))^{-1}\partial_\mu S_I^L(x;1)=\sum_a\mathcal{A}^a_\mu(x)t^a_L,
\\
&(P^L(x;1,I))^{-1}\partial_\mu P^L(x;1,I)=\sum_a\mathcal{B}^a_\mu(x)t^a_L .
\end{align}
In deriving the cubic term below, we use
\begin{align}
S\partial_\mu S^{-1}
=
-S(S^{-1}\partial_\mu S)S^{-1},
\end{align}
together with the cyclicity of the trace. This gives one overall minus sign in
the cubic trace. Then, as in the corresponding continuum calculation, we find
\begin{align}
&\frac{1}{12\pi}\sum_{c\in\Lambda_3}
    \int_c d^3x \sum_{\mu,\nu,\rho}
    \epsilon_{\mu\nu\rho}
    \Tr\{
      S_I^L(x;1)\partial_\mu (S_I^L(x;1))^{-1}
      \,S_I^L(x;1)\partial_\nu (S_I^L(x;1))^{-1}
      \,S_I^L(x;1)\,\partial_\rho (S_I^L(x;1))^{-1}
    \}
    \notag\\
    &\qquad
    +
    \frac{1}{4\pi}\sum_{c\in\Lambda_3}\sum_{p\subset\partial c}
    \int_p d^2x\,
    \sum_{\mu,\nu,\rho}
    n_p^\mu\,
    \epsilon_{\mu\nu\rho}\,
    \Tr\{
      (P^L(x;1,I))^{-1}\partial_\nu P^L(x;1,I)
      \,(S_I^L(x;1))^{-1}\partial_\rho S_I^L(x;1)
    \}\notag\\
    &=-\frac{1}{12\pi}\sum_{c\in\Lambda_3}
    \int_c d^3x \sum_{\mu,\nu,\rho}
    \epsilon_{\mu\nu\rho}
    \sum_{a,b,e}\mathcal{A}^a_\mu(x)\mathcal{A}^b_\nu(x)\mathcal{A}^e_\rho(x)\Tr\{
      t^a_Lt^b_Lt^e_L
    \}
    \notag\\
    &\qquad
    +
    \frac{1}{4\pi}\sum_{c\in\Lambda_3}\sum_{p\subset\partial c}
    \int_p d^2x\,
    \sum_{\mu,\nu,\rho}
    n_p^\mu\,
    \epsilon_{\mu\nu\rho}\,
    \sum_{a,b}\mathcal{B}^a_\nu(x)\mathcal{A}^b_\rho(x)\Tr\{t^a_Lt^b_L\}
    \notag\\
    &=-\frac{1}{24\pi}\sum_{c\in\Lambda_3}
    \int_c d^3x \sum_{\mu,\nu,\rho}
    \epsilon_{\mu\nu\rho}
    \sum_{a,b,e}\im\mathcal{A}^a_\mu(x)\mathcal{A}^b_\nu(x)\mathcal{A}^e_\rho(x)
    \sum_{d}f^{bed}\Tr\{
      t^a_Lt^d_L
    \}
    \notag\\
    &\qquad
    +
    \frac{1}{4\pi}\sum_{c\in\Lambda_3}\sum_{p\subset\partial c}
    \int_p d^2x\,
    \sum_{\mu,\nu,\rho}
    n_p^\mu\,
    \epsilon_{\mu\nu\rho}\,
    \sum_{a,b}\mathcal{B}^a_\nu(x)\mathcal{A}^b_\rho(x)\Tr\{t^a_Lt^b_L\}.
    \label{rep dependence}
\end{align}
The representation dependence of the cubic trace reduces to the quadratic form
because, due to the antisymmetry of $\epsilon_{\mu\nu\rho}$, only the
antisymmetric part in the indices $b$ and $e$ contributes. We can then use
\begin{align}
[t_L^b,t_L^e]=\im\sum_d f^{bed}t_L^d .
\end{align}

The right-sector contribution $\sum_{c\in\Lambda_3} k_c^R[g_2^\dagger,U]$ is
evaluated by applying the replacement
\begin{align}
    R_L\to R_R,
    \qquad
    g_1\to g_2^\dagger
\end{align}
to the preceding argument. We define
\begin{align}
    [g_2^c(x)]^\dagger
    &:=
    \{S^R_I(x;1)\}^{-1}g_2(0)^\dagger S^R(x;g_2^\dagger),
    \label{def smooth g_2}
    \\
    S^R(x;g_2^\dagger)
    &=
    g_2(0) S^R_I(x;1)[g_2^c(x)]^\dagger
    \label{gauge tr S^R}
\end{align}
in the same way as Eq.~\eqref{def smooth g_1}. Then the right-sector contribution
is decomposed as
\begin{align}
    &\sum_{c\in\Lambda_3}k_c^R[g_2^\dagger,U]\notag\\
    &=\frac{1}{12\pi}\sum_{c\in\Lambda_3}
    \int_c d^3x \sum_{\mu,\nu,\rho}
    \epsilon_{\mu\nu\rho}
    \Tr\{
      g_2^c(x)\,\partial_\mu  (g_2^c(x))^{-1}
      \,g_2^c(x)\,\partial_\nu  (g_2^c(x))^{-1}
      \,g_2^c(x)\,\partial_\rho  (g_2^c(x))^{-1}
    \}
    \notag\\
    &
    +
    \frac{1}{4\pi}\sum_{c\in\Lambda_3}\sum_{p\subset\partial c}
    \int_p d^2x\,
    \sum_{\mu,\nu,\rho}
    n_p^\mu\,
    \epsilon_{\mu\nu\rho}\,
    \Tr\{
      (P^R(x;1,I))^{-1}\partial_\nu P^R(x;1,I)
    \,g_2^c(x)\,\partial_\rho  (g_2^c(x))^{-1}
    \}\notag\\
    &+\frac{1}{12\pi}\sum_{c\in\Lambda_3}
    \int_c d^3x \sum_{\mu,\nu,\rho}
    \epsilon_{\mu\nu\rho}
    \Tr\{
      S_I^R(x;1)\partial_\mu (S_I^R(x;1))^{-1}
      \,S_I^R(x;1)\partial_\nu (S_I^R(x;1))^{-1}
      \,S_I^R(x;1)\,\partial_\rho (S_I^R(x;1))^{-1}
    \}
    \notag\\
    &
    +
   \frac{1}{4\pi}\sum_{c\in\Lambda_3}\sum_{p\subset\partial c}
    \int_p d^2x\,
    \sum_{\mu,\nu,\rho}
    n_p^\mu\,
    \epsilon_{\mu\nu\rho}\,
    \Tr\{
      (P^R(x;1,I))^{-1}\partial_\nu P^R(x;1,I)
      \,(S_I^R(x;1))^{-1}\partial_\rho S_I^R(x;1)
    \}.
   \label{right k_c decomposed}
\end{align}
The first two terms depend only on the boundary degrees of freedom by the same
argument as that leading to Eqs.~\eqref{lattice WZ} and
\eqref{left boundary contribution}.

In the right sector, the interpolation is also defined by mapping the
same fundamental-representation interpolation to the right representation by
$R_R$. Therefore, with the same coefficients
$\mathcal A^a_\mu(x)$ and $\mathcal B^a_\mu(x)$ as in the left sector, we can write
\begin{align}
&(S_I^R(x;1))^{-1}\partial_\mu S_I^R(x;1)=\sum_a\mathcal{A}^a_\mu(x)t^a_R,
\\
&(P^R(x;1,I))^{-1}\partial_\mu P^R(x;1,I)=\sum_a\mathcal{B}^a_\mu(x)t^a_R.
\end{align}
For the second part of Eq.~\eqref{right k_c decomposed}, the only difference from
Eq.~\eqref{rep dependence} is the choice of representation. We therefore obtain
\begin{align}
&\frac{1}{12\pi}\sum_{c\in\Lambda_3}
    \int_c d^3x \sum_{\mu,\nu,\rho}
    \epsilon_{\mu\nu\rho}
    \Tr\{
      S_I^R(x;1)\partial_\mu (S_I^R(x;1))^{-1}
      \,S_I^R(x;1)\partial_\nu (S_I^R(x;1))^{-1}
      \,S_I^R(x;1)\,\partial_\rho (S_I^R(x;1))^{-1}
    \}
    \notag\\
    &\qquad
    +
    \frac{1}{4\pi}\sum_{c\in\Lambda_3}\sum_{p\subset\partial c}
    \int_p d^2x\,
    \sum_{\mu,\nu,\rho}
    n_p^\mu\,
    \epsilon_{\mu\nu\rho}\,
    \Tr\{
      (P^R(x;1,I))^{-1}\partial_\nu P^R(x;1,I)
      \,(S_I^R(x;1))^{-1}\partial_\rho S_I^R(x;1)
    \}\notag\\
    &=-\frac{1}{12\pi}\sum_{c\in\Lambda_3}
    \int_c d^3x \sum_{\mu,\nu,\rho}
    \epsilon_{\mu\nu\rho}
    \sum_{a,b,e}\mathcal{A}^a_\mu(x)\mathcal{A}^b_\nu(x)\mathcal{A}^e_\rho(x)\Tr\{
      t^a_Rt^b_Rt^e_R
    \}
    \notag\\
    &\qquad
    +
    \frac{1}{4\pi}\sum_{c\in\Lambda_3}\sum_{p\subset\partial c}
    \int_p d^2x\,
    \sum_{\mu,\nu,\rho}
    n_p^\mu\,
    \epsilon_{\mu\nu\rho}\,
    \sum_{a,b}\mathcal{B}^a_\nu(x)\mathcal{A}^b_\rho(x)\Tr\{t^a_Rt^b_R\}
    \notag\\
    &=-\frac{1}{24\pi}\sum_{c\in\Lambda_3}
    \int_c d^3x \sum_{\mu,\nu,\rho}
    \epsilon_{\mu\nu\rho}
    \sum_{a,b,e}\im\mathcal{A}^a_\mu(x)\mathcal{A}^b_\nu(x)\mathcal{A}^e_\rho(x)
    \sum_{d}f^{bed}\Tr\{
      t^a_Rt^d_R
    \}
    \notag\\
    &\qquad
    +
    \frac{1}{4\pi}\sum_{c\in\Lambda_3}\sum_{p\subset\partial c}
    \int_p d^2x\,
    \sum_{\mu,\nu,\rho}
    n_p^\mu\,
    \epsilon_{\mu\nu\rho}\,
    \sum_{a,b}\mathcal{B}^a_\nu(x)\mathcal{A}^b_\rho(x)\Tr\{t^a_Rt^b_R\}.
    \label{rep dependence right}
\end{align}

Therefore, if the representations of the gauge group satisfy
\begin{align}
\Tr\{t^a_Lt^b_L\}
=
\Tr\{t^a_Rt^b_R\},
\end{align}
then Eqs.~\eqref{rep dependence} and~\eqref{rep dependence right} cancel each
other in the factor
\begin{align}
    \exp\left(
    \im\sum_{c\in\Lambda_3} k_c^L[g_1,U]
    -
    \im\sum_{c\in\Lambda_3} k_c^R[g_2^\dagger,U]
    \right).
\end{align}
Explicitly,
\begin{align}
&-\frac{1}{24\pi}\sum_{c\in\Lambda_3}
    \int_c d^3x \sum_{\mu,\nu,\rho}
    \epsilon_{\mu\nu\rho}
    \sum_{a,b,e}\im\mathcal{A}^a_\mu(x)\mathcal{A}^b_\nu(x)\mathcal{A}^e_\rho(x)
    \sum_{d}f^{bed} \left[
    \Tr\{
      t^a_Lt^d_L
    \}-\Tr\{
      t^a_Rt^d_R
    \}
    \right]
    \notag\\
    &\qquad
    +
    \frac{1}{4\pi}\sum_{c\in\Lambda_3}\sum_{p\subset\partial c}
    \int_p d^2x\,
    \sum_{\mu,\nu,\rho}
    n_p^\mu\,
    \epsilon_{\mu\nu\rho}\,
    \sum_{a,b}\mathcal{B}^a_\nu(x)\mathcal{A}^b_\rho(x) \left[
    \Tr\{
      t^a_Lt^b_L
    \}-\Tr\{
      t^a_Rt^b_R
    \}
    \right]=0.
\end{align}

Thus, when the anomaly-cancellation condition is satisfied, the bulk contribution
coming from the pure gauge field cancels between the left and right sectors. The
remaining terms are the ordinary Wess--Zumino terms and boundary-localized
terms. Hence the Boltzmann factor is independent of the bulk extension, up to the standard
integer ambiguity of the ordinary Wess--Zumino terms, and is determined by the
boundary degrees of freedom. Namely,
\begin{align}
&\exp\left(
\im\sum_{c\in\Lambda_3} k_c^L[g_1,U]
-\im\sum_{c\in\Lambda_3} k_c^R[g_2^\dagger,U]
\right)
\notag\\
&=
\exp
\left[
\frac{\im}{12\pi}
    \int_{M_3} d^3\xi\sum_{\mu,\nu,\rho}
    \epsilon_{\mu\nu\rho}
    \Tr\{
      (g_1(\xi))^{-1}\,\partial_\mu g_1(\xi)
      \,(g_1(\xi))^{-1}\,\partial_\nu g_1(\xi)
      \,(g_1(\xi))^{-1}\,\partial_\rho g_1(\xi)
    \}
\right.
\notag\\
&\qquad\left.
+\frac{\im}{4\pi}\sum_{p\subset\Lambda_2}
    \int_p d^2x\,
    \sum_{\mu,\nu,\rho}
    n_p^\mu\,
    \epsilon_{\mu\nu\rho}\,
    \Tr\{
      (P^L(x;1,I))^{-1}\partial_\nu P^L(x;1,I)
    \,(g_1^c(x))^{-1}\partial_\rho g_1^c(x)
    \}\right.
\notag\\
&\qquad\left.
-\frac{\im}{12\pi}
    \int_{M_3} d^3\xi\sum_{\mu,\nu,\rho}
    \epsilon_{\mu\nu\rho}
    \Tr\{
      g_2(\xi)\,\partial_\mu (g_2(\xi))^{-1}
      \,g_2(\xi)\,\partial_\nu (g_2(\xi))^{-1}
      \,g_2(\xi)\,\partial_\rho (g_2(\xi))^{-1}
    \}\right.
\notag\\
&\qquad\left.
-\frac{\im}{4\pi}\sum_{p\subset\Lambda_2}
    \int_p d^2x\,
    \sum_{\mu,\nu,\rho}
    n_p^\mu\,
    \epsilon_{\mu\nu\rho}\,
    \Tr\{
      (P^R(x;1,I))^{-1}\partial_\nu P^R(x;1,I)
    \,g_2^c(x)\,\partial_\rho (g_2^c(x))^{-1}
    \}
\right].
\end{align}
The $M_3$ integrals on the right-hand side are ordinary Wess--Zumino terms, and
their exponentials are independent of the choice of extension. Therefore, under
the anomaly-free condition, the bulk dependence originating from the background
gauge field disappears, and the Boltzmann factor of the lattice action can be
described only in terms of boundary two-dimensional degrees of freedom.

We have thus shown that the Boltzmann factor $e^{-S[g_1,g_2,U]}$ defined by the
gauge-invariant lattice action
\begin{align}
    S[g_1,g_2,U]
    &=
    \sum_{n\in\Lambda_2}
    \frac{1}{4\lambda^2}
    \sum_\mu
    \Tr\left[\bigl(D^L g_1(n,\mu)\bigr)^\dagger\bigl(D^L g_1(n,\mu)\bigr)\right]
    \notag\\
    &\qquad+
    \sum_{n\in\Lambda_2}
    \frac{1}{4\lambda^2}
    \sum_\mu
    \Tr\left[\bigl(D^R g_2(n,\mu)\bigr)^\dagger\bigl(D^R g_2(n,\mu)\bigr)\right]
    \notag\\
    &\qquad\qquad+
   \im \sum_{c\in\Lambda_3} k_c^L[g_1,U]
    -
    \im\sum_{c\in\Lambda_3} k_c^R[g_2^\dagger,U]
\end{align}
can be written using only terms depending on degrees of freedom on
$\Lambda_2=\partial\Lambda_3$, provided that the representations of the gauge
group satisfy
\begin{align}
\Tr\{t^a_Lt^b_L\}
=
\Tr\{t^a_Rt^b_R\}
\qquad
\text{for all }a,b.
\end{align}
This is the finite-lattice anomaly-cancellation mechanism in the bosonized
lattice formulation of the two-dimensional non-Abelian chiral gauge theory.

\section{A strategy to extract observables}
\label{sec:observables}

In this section, we describe a prescription for extracting correlation functions
of the original fermion theory from the bosonized theory with spectator fermions.
The prescription given below assumes that, after a suitable fine tuning, the continuum
limit of the lattice model flows to the gauged WZW model which is equivalent,
via non-Abelian bosonization, to the Dirac fermion theory including the spectator
fermions. In particular, we assume that the corresponding bosonization dictionary is
valid in this continuum limit.

The correlation functions considered in this section are those of operators
inserted on the two-dimensional boundary lattice. They are evaluated with the
gauge field also integrated over, under the anomaly-cancellation condition.
We assume, however, that the gauge-field path integral is well-defined. In particular, the
lattice Chern--Simons/WZW-type bulk term appearing in our construction may
leave a quantum-mechanical subtlety associated with extra zero modes
\cite{Berruto:2000dp,Berruto:2000fb}. We return to this point in the Conclusion.
In the present section, we formulate the observable extraction prescription under
the assumption that such a potential zero-mode problem is appropriately
controlled and that the lattice theory can define correlation functions.

For observables in the original fermion theory, we use the factorization
explained around Eq.~\eqref{aux=original}. The spectator fermions are free and
decoupled. Hence they do not affect normalized correlation functions of
operators that contain no spectator fields.

In particular, Wilson loops contain no spectator
fermions. Their expectation values therefore agree with those in the original
fermion theory.

Thus, the nontrivial observables to be discussed below are fermion bilinears
connected by open Wilson lines and their correlation functions. These bilocal operators provide
gauge-covariant point-splittings of local fermion bilinears. When the two
endpoints are brought close to each other, their short-distance expansion contains
flavor-contracted local operators such as
$\Psi_{L/R}^\dagger\Psi_{L/R}$ and
$\Psi_{L/R}^\dagger D_\mu^{L/R}\Psi_{L/R}$.
We therefore take open-Wilson-line bilinears as the basic class of operators.

The basic idea is not to represent a single fermion operator itself as a local
bosonic variable. Instead, we first bosonize bilinear operators which contain the
spectator fermions, and then divide out the correlation function of the free
spectator sector. In this way, we extract the information of the original fermion sector.

To make this point explicit, recall that the dynamical bosonic variables in our
action are $g_1$ and $g_2$. The non-Abelian bosonization dictionary~\cite{Witten:nonabelian-bosonization} is
\begin{align}
    &M_1 (g_1)_{ij}
    =
    -\im \psi_{L,i}\psi^{\text{sp}}_{R,j},
    &
    M_2 (g_2)_{ij}
    =
    -\im \psi^{\text{sp}}_{L,i}\psi_{R,j}.\label{dictionary}
\end{align}
Here $M_1$ and $M_2$ have mass dimension one and depend on the choice of
normalization and renormalization prescription.

Therefore, correlation functions constructed only from $g_1$ and $g_2$
necessarily contain the spectator fermion sector. In non-Abelian bosonization,
fermion bilinears have simple representations in the bosonization dictionary~\cite{Witten:nonabelian-bosonization}.

By contrast, a single fermion operator is more indirect. In the continuum CFT,
it can be identified with a product of a vertex operator in the $U(1)$ sector and
a primary field in the non-Abelian sector~\cite{james2018non}. This is an
operator identification in the continuum CFT. It is not a local bosonic variable
constructed directly from the lattice fields $g_1$ and $g_2$.

We therefore do not attempt to implement a single fermion operator at finite
lattice spacing. Instead, we use bilinear operators involving spectator fermions.

The present prescription uses the fact that the spectator fermions form a
decoupled free sector. We explicitly factor out the contribution of the spectator
sector from correlation functions of bilinear operators involving spectator
fermions. This gives a way to extract gauge-invariant multipoint correlation
functions in the original fermion sector from the bosonized theory, without
requiring a direct lattice realization of a single fermion operator.

The strategy is as follows. Consider the following gauge-invariant multipoint
function in the original fermion theory:
\begin{align}
\left\langle
\prod_{a=1}^{n_L}
\left(
\sum_{i,j}
\psi_{L,i}^\dagger(x_a)
\mathcal{W}_{ij}^L(x_a,x_a')
\psi_{L,j}(x_a')
\right)
\prod_{b=1}^{n_R}
\left(
\sum_{k,l}
\psi_{R,k}^\dagger(y_b)
\mathcal{W}_{kl}^R(y_b,y_b')
\psi_{R,l}(y_b')
\right)
\right\rangle .
\end{align}
Here $\mathcal{W}_{ij}^{L/R}(x_a,x_a')$ is defined by
\begin{align}
\mathcal{W}_{ij}^{L/R}(x_a,x_a')
:=
\left(
\mathcal{P}_{x_a\to x_a'}
e^{\im\int_{\text{path}}A^{L/R}}
\right)_{ij}.
\end{align}
The symbol $\mathcal{P}_{x_a\to x_a'}\exp{}$ denotes the path-ordered
exponential along a chosen path connecting the spacetime points $x_a$ and
$x_a'$.

We now dress this correlation function only by spectator fermions:
\begin{align}
&\left\langle
\prod_{a=1}^{n_L}
\left(
\sum_{i,j}\sum_m
{\psi_{R,m}^{\text{sp}}}^\dagger(x_a)
\psi_{L,i}^\dagger(x_a)
\mathcal{W}_{ij}^L(x_a,x_a')
\psi_{L,j}(x_a')
\psi_{R,m}^{\text{sp}}(x_a')
\right)
\right.
\notag\\
&\qquad\qquad\qquad
\times
\left.
\prod_{b=1}^{n_R}
\left(
\sum_{k,l}\sum_n
\psi_{R,k}^\dagger(y_b)
{\psi_{L,n}^{\text{sp}}}^\dagger(y_b)
\mathcal{W}_{kl}^R(y_b,y_b')
\psi_{L,n}^{\text{sp}}(y_b')
\psi_{R,l}(y_b')
\right)
\right\rangle .
\end{align}
Since the spectator fermions do not interact with the original fermions, this
dressed correlation function decomposes as
\begin{align}
&\left\langle
\prod_{a=1}^{n_L}
\left(
\sum_{i,j}\sum_m
{\psi_{R,m}^{\text{sp}}}^\dagger(x_a)
\psi_{L,i}^\dagger(x_a)
\mathcal{W}_{ij}^L(x_a,x_a')
\psi_{L,j}(x_a')
\psi_{R,m}^{\text{sp}}(x_a')
\right)
\right.
\notag\\
&\qquad\qquad\qquad
\times
\left.
\prod_{b=1}^{n_R}
\left(
\sum_{k,l}\sum_n
\psi_{R,k}^\dagger(y_b)
{\psi_{L,n}^{\text{sp}}}^\dagger(y_b)
\mathcal{W}_{kl}^R(y_b,y_b')
\psi_{L,n}^{\text{sp}}(y_b')
\psi_{R,l}(y_b')
\right)
\right\rangle
\notag\\
&=
\left\langle
\prod_{a=1}^{n_L}
\left(
\sum_m
{\psi_{R,m}^{\text{sp}}}^\dagger(x_a)
\psi_{R,m}^{\text{sp}}(x_a')
\right)
\prod_{b=1}^{n_R}
\left(
\sum_n
{\psi_{L,n}^{\text{sp}}}^\dagger(y_b)
\psi_{L,n}^{\text{sp}}(y_b')
\right)
\right\rangle
\notag\\
&\quad\times
\left\langle
\prod_{a=1}^{n_L}
\left(
\sum_{i,j}
\psi_{L,i}^\dagger(x_a)
\mathcal{W}_{ij}^L(x_a,x_a')
\psi_{L,j}(x_a')
\right)
\prod_{b=1}^{n_R}
\left(
\sum_{k,l}
\psi_{R,k}^\dagger(y_b)
\mathcal{W}_{kl}^R(y_b,y_b')
\psi_{R,l}(y_b')
\right)
\right\rangle .
\end{align}
On the other hand, by applying the non-Abelian bosonization dictionary
\eqref{dictionary} and its Hermitian conjugate, the same dressed correlation
function is mapped to
\begin{align}
&\left\langle
\prod_{a=1}^{n_L}
\left(
\sum_{i,j}\sum_m
{\psi_{R,m}^{\text{sp}}}^\dagger(x_a)
\psi_{L,i}^\dagger(x_a)
\mathcal{W}_{ij}^L(x_a,x_a')
\psi_{L,j}(x_a')
\psi_{R,m}^{\text{sp}}(x_a')
\right)
\right.
\notag\\
&\qquad\qquad\qquad
\times
\left.
\prod_{b=1}^{n_R}
\left(
\sum_{k,l}\sum_n
\psi_{R,k}^\dagger(y_b)
{\psi_{L,n}^{\text{sp}}}^\dagger(y_b)
\mathcal{W}_{kl}^R(y_b,y_b')
\psi_{L,n}^{\text{sp}}(y_b')
\psi_{R,l}(y_b')
\right)
\right\rangle
\notag\\
&=
(M_1)^{2n_L}(M_2)^{2n_R}
\left\langle
\prod_{a=1}^{n_L}
\Tr\{
g_1^\dagger(x_a)
\mathcal{W}^L(x_a,x_a')
g_1(x_a')
\}
\prod_{b=1}^{n_R}
\Tr\{
g_2^\dagger(y_b)
\mathcal{W}^R(y_b,y_b')
g_2(y_b')
\}
\right\rangle .
\end{align}
Here we introduce
\begin{align}
\mathcal{W}^{L/R}(x_a,x_a')
:=
\mathcal{P}_{x_a\to x_a'}
e^{\im\int_{\text{path}}A^{L/R}} .
\end{align}
Combining the two expressions, we obtain
\begin{align}
&(M_1)^{2n_L}(M_2)^{2n_R}
\left\langle
\prod_{a=1}^{n_L}
\Tr\{
g_1^\dagger(x_a)
\mathcal{W}^L(x_a,x_a')
g_1(x_a')
\}
\prod_{b=1}^{n_R}
\Tr\{
g_2^\dagger(y_b)
\mathcal{W}^R(y_b,y_b')
g_2(y_b')
\}
\right\rangle
\notag\\
&=
\left\langle
\prod_{a=1}^{n_L}
\left(
\sum_m
{\psi_{R,m}^{\text{sp}}}^\dagger(x_a)
\psi_{R,m}^{\text{sp}}(x_a')
\right)
\prod_{b=1}^{n_R}
\left(
\sum_n
{\psi_{L,n}^{\text{sp}}}^\dagger(y_b)
\psi_{L,n}^{\text{sp}}(y_b')
\right)
\right\rangle
\notag\\
&\quad\times
\left\langle
\prod_{a=1}^{n_L}
\left(
\sum_{i,j}
\psi_{L,i}^\dagger(x_a)
\mathcal{W}_{ij}^L(x_a,x_a')
\psi_{L,j}(x_a')
\right)
\prod_{b=1}^{n_R}
\left(
\sum_{k,l}
\psi_{R,k}^\dagger(y_b)
\mathcal{W}_{kl}^R(y_b,y_b')
\psi_{R,l}(y_b')
\right)
\right\rangle .
\end{align}
Therefore, if the spectator prefactor satisfies
\begin{align}
\left\langle
\prod_{a=1}^{n_L}
\left(
\sum_m
{\psi_{R,m}^{\text{sp}}}^\dagger(x_a)
\psi_{R,m}^{\text{sp}}(x_a')
\right)
\prod_{b=1}^{n_R}
\left(
\sum_n
{\psi_{L,n}^{\text{sp}}}^\dagger(y_b)
\psi_{L,n}^{\text{sp}}(y_b')
\right)
\right\rangle
\neq0,
\label{aux fermion=0}
\end{align}
we obtain
\begin{align}
&\left\langle
\prod_{a=1}^{n_L}
\left(
\sum_{i,j}
\psi_{L,i}^\dagger(x_a)
\mathcal{W}_{ij}^L(x_a,x_a')
\psi_{L,j}(x_a')
\right)
\prod_{b=1}^{n_R}
\left(
\sum_{k,l}
\psi_{R,k}^\dagger(y_b)
\mathcal{W}_{kl}^R(y_b,y_b')
\psi_{R,l}(y_b')
\right)
\right\rangle
\notag\\
&=
\frac{
(M_1)^{2n_L}(M_2)^{2n_R}
\left\langle
\prod_{a=1}^{n_L}
\Tr\{
g_1^\dagger(x_a)
\mathcal{W}^L(x_a,x_a')
g_1(x_a')
\}
\prod_{b=1}^{n_R}
\Tr\{
g_2^\dagger(y_b)
\mathcal{W}^R(y_b,y_b')
g_2(y_b')
\}
\right\rangle
}{
\left\langle
\prod_{a=1}^{n_L}
\left(
\sum_m
{\psi_{R,m}^{\text{sp}}}^\dagger(x_a)
\psi_{R,m}^{\text{sp}}(x_a')
\right)
\prod_{b=1}^{n_R}
\left(
\sum_n
{\psi_{L,n}^{\text{sp}}}^\dagger(y_b)
\psi_{L,n}^{\text{sp}}(y_b')
\right)
\right\rangle
}.
\label{boson/spectator}
\end{align}
This formula expresses correlation functions of the original chiral gauge theory
in terms of correlation functions in the bosonized theory and in the spectator
free Dirac fermion theory.

It remains to check when the denominator in Eq.~\eqref{boson/spectator} is
nonzero. The spectator factor is a singlet under the global
$U(N_f)_L\times U(N_f)_R$ symmetry of the massless free Dirac fermion. Therefore,
it is not forced to vanish by a selection rule associated with this symmetry.
More explicitly, with the convention
\begin{align}
    \left\langle
    {\psi_{R,m}^{\text{sp}}}^\dagger(x)
    \psi_{R,n}^{\text{sp}}(x')
    \right\rangle
    =
    \frac{\delta_{mn}}{2\pi}
    \frac{1}{z-z'},
    \qquad
    \left\langle
    {\psi_{L,m}^{\text{sp}}}^\dagger(y)
    \psi_{L,n}^{\text{sp}}(y')
    \right\rangle
    =
    \frac{\delta_{mn}}{2\pi}
    \frac{1}{\bar w-\bar w'},
\end{align}
where
\begin{align}
&z=x^1+\im x^2,
\qquad
\bar z=x^1-\im x^2,
\qquad
w=y^1+\im y^2,
\qquad
\bar w=y^1-\im y^2,
\end{align}
one obtains
\begin{align}
&\left\langle
\prod_{a=1}^{n_L}
\left(
\sum_m
{\psi_{R,m}^{\text{sp}}}^\dagger(x_a)
\psi_{R,m}^{\text{sp}}(x_a')
\right)
\prod_{b=1}^{n_R}
\left(
\sum_n
{\psi_{L,n}^{\text{sp}}}^\dagger(y_b)
\psi_{L,n}^{\text{sp}}(y_b')
\right)
\right\rangle
\notag\\
&\qquad =
\left(\frac{1}{2\pi}\right)^{n_L+n_R}
\left[
\sum_{\sigma\in S_{n_L}}
\text{sgn}(\sigma)
N_f^{c(\sigma)}
\prod_{a=1}^{n_L}
\frac{1}{z_a-z'_{\sigma(a)}}
\right]
\left[
\sum_{\tau\in S_{n_R}}
\text{sgn}(\tau)
N_f^{c(\tau)}
\prod_{b=1}^{n_R}
\frac{1}{\bar w_b-\bar w'_{\tau(b)}}
\right].
\label{spectator corelation continuum}
\end{align}
Here $S_n$ is the symmetric group of degree $n$, and $c(\sigma)$ is the number
of disjoint cycles in the permutation $\sigma$. For a fixed Wick contraction
specified by $\sigma$, the flavor sum gives
\begin{align}
\sum_{m_1,\ldots,m_n}
\prod_{a=1}^{n}
\delta_{m_a m_{\sigma(a)}}
=
N_f^{c(\sigma)}.
\end{align}
Indeed, the Kronecker deltas identify all flavor indices belonging to the same
cycle of $\sigma$. Each cycle leaves one independent flavor sum and therefore
gives one factor of $N_f$.

Thus, the denominator in Eq.~\eqref{boson/spectator} is not forced to vanish by
a global-symmetry selection rule. For generic insertion points, it is nonzero.
Possible zeros, if any, can occur only for special choices of insertion points,
where different Wick contractions cancel. We apply Eq.~\eqref{boson/spectator}
away from such exceptional choices.

\paragraph{Lattice to continuum}

We finally describe how the above prescription is implemented starting from the
lattice formulation. First, the continuum parallel transporter
$\mathcal{W}^{L/R}(x_a,x_a')$ is replaced by the product of link variables along
a path on the lattice. The corresponding lattice version of the bosonic correlation
function is
\begin{align}
&\left\langle
\prod_{a=1}^{n_L}
\Tr\{
g_1^\dagger(x_a)
\mathcal{W}^L(x_a,x_a')
g_1(x_a')
\}
\prod_{b=1}^{n_R}
\Tr\{
g_2^\dagger(y_b)
\mathcal{W}^R(y_b,y_b')
g_2(y_b')
\}
\right\rangle
\notag\\
&\to
\left\langle
\prod_{a=1}^{n_L}
\Tr\left\{
g_1^\dagger(n_a)
\left(
\prod_{(nm)\in\text{Path}:n_a\to n_a'}
U^L_{nm}
\right)
g_1(n_a')
\right\}
\right.
\notag\\
&\qquad\qquad\qquad
\times
\left.
\prod_{b=1}^{n_R}
\Tr\left\{
g_2^\dagger(m_b)
\left(
\prod_{(pq)\in\text{Path}:m_b\to m_b'}
U^R_{pq}
\right)
g_2(m_b')
\right\}
\right\rangle_{\text{lattice}},
\notag\\
&\qquad\qquad
n_a,\,n_a',\,m_b,\,m_b'\in\Lambda_2 .
\end{align}
The denominator is similarly replaced by the corresponding lattice spectator
correlator,
\begin{align}
&\left\langle
\prod_{a=1}^{n_L}
\left(
\sum_m
{\psi_{R,m}^{\text{sp}}}^\dagger(x_a)
\psi_{R,m}^{\text{sp}}(x_a')
\right)
\prod_{b=1}^{n_R}
\left(
\sum_n
{\psi_{L,n}^{\text{sp}}}^\dagger(y_b)
\psi_{L,n}^{\text{sp}}(y_b')
\right)
\right\rangle
\notag\\
&\to
\left\langle
\prod_{a=1}^{n_L}
\left(
\sum_m
{\psi_{R,m}^{\text{sp}}}^\dagger(n_a)
\psi_{R,m}^{\text{sp}}(n_a')
\right)
\prod_{b=1}^{n_R}
\left(
\sum_n
{\psi_{L,n}^{\text{sp}}}^\dagger(m_b)
\psi_{L,n}^{\text{sp}}(m_b')
\right)
\right\rangle_{\text{lattice}} .
\end{align}
The choice of the lattice action for the spectator fermions requires some care.
A naive discretization produces doublers. The spectator fermions enter only
through the denominator of Eq.~\eqref{boson/spectator}, which is a correlator of
a free massless Dirac fermion theory. Thus, on the lattice, the spectator action
is only required to reproduce the desired free spectator correlator in the
continuum limit. If a naive discretization is used, the extra low-energy
contribution from doublers must be removed. Alternatively, one may use a
regularization in which the doubler modes are lifted. The only requirement is
that the lattice spectator correlator has the correct continuum limit, namely
\begin{align}
&\left\langle
\prod_{a=1}^{n_L}
\left(
\sum_m
{\psi_{R,m}^{\text{sp}}}^\dagger(n_a)
\psi_{R,m}^{\text{sp}}(n_a')
\right)
\prod_{b=1}^{n_R}
\left(
\sum_n
{\psi_{L,n}^{\text{sp}}}^\dagger(m_b)
\psi_{L,n}^{\text{sp}}(m_b')
\right)
\right\rangle_{\text{lattice}}
\notag\\
&\xrightarrow{\text{continuum limit}}
\left(\frac{1}{2\pi}\right)^{n_L+n_R}
\left[
\sum_{\sigma\in S_{n_L}}
\text{sgn}(\sigma)
N_f^{c(\sigma)}
\prod_{a=1}^{n_L}
\frac{1}{z_a-z'_{\sigma(a)}}
\right]
\notag\\
&\qquad\qquad\qquad\qquad\qquad
\times
\left[
\sum_{\tau\in S_{n_R}}
\text{sgn}(\tau)
N_f^{c(\tau)}
\prod_{b=1}^{n_R}
\frac{1}{\bar w_b-\bar w'_{\tau(b)}}
\right].
\end{align}
Under this condition, the correlation function of the original chiral gauge theory
is obtained in the continuum limit as
\begin{align}
&\left\langle
\prod_{a=1}^{n_L}
\Tr\left\{
g_1^\dagger(n_a)
\left(
\prod_{(nm)\in\text{Path}:n_a\to n_a'}
U^L_{nm}
\right)
g_1(n_a')
\right\}
\prod_{b=1}^{n_R}
\Tr\left\{
g_2^\dagger(m_b)
\left(
\prod_{(pq)\in\text{Path}:m_b\to m_b'}
U^R_{pq}
\right)
g_2(m_b')
\right\}
\right\rangle_{\text{lattice}}
\notag\\
&\qquad\qquad\qquad\times
\frac{(M_1)^{2n_L}(M_2)^{2n_R}}
{
\left\langle
\prod_{a=1}^{n_L}
\left(
\sum_m
{\psi_{R,m}^{\text{sp}}}^\dagger(n_a)
\psi_{R,m}^{\text{sp}}(n_a')
\right)
\prod_{b=1}^{n_R}
\left(
\sum_n
{\psi_{L,n}^{\text{sp}}}^\dagger(m_b)
\psi_{L,n}^{\text{sp}}(m_b')
\right)
\right\rangle_{\text{lattice}}
}
\notag\\
&\xrightarrow{\text{continuum limit}}
\left\langle
\prod_{a=1}^{n_L}
\left(
\sum_{i,j}
\psi_{L,i}^\dagger(x_a)
\mathcal{W}_{ij}^L(x_a,x_a')
\psi_{L,j}(x_a')
\right)
\prod_{b=1}^{n_R}
\left(
\sum_{k,l}
\psi_{R,k}^\dagger(y_b)
\mathcal{W}_{kl}^R(y_b,y_b')
\psi_{R,l}(y_b')
\right)
\right\rangle .
\end{align}

\section{Conclusion and discussion}
\label{sec:conclusion}

In this paper, we have proposed a lattice formulation of a bosonic theory obtained
via non-Abelian bosonization. The fermionic theory we aim to describe is a
two-dimensional non-Abelian chiral gauge theory. After adding gauge-neutral
spectator fermions, the theory can be reorganized into two one-sided gauged WZW sectors. 
We constructed a lattice action for this reorganized theory.

The action consists of two parts. The first part is the kinetic term on the
boundary $\Lambda_2=\partial\Lambda_3$. The second part is the
Wess--Zumino/Chern--Simons-type term in the three-dimensional bulk~$\Lambda_3$.
We showed that the left and right bulk contributions cancel in the exponentiated
action under the anomaly-free condition
\begin{align}
    \Tr\left(t^a_Lt^b_L\right)
    =
    \Tr\left(t^a_Rt^b_R\right)
    \qquad
    \text{for all } a,b.
\end{align}
Thus, the Boltzmann factor reduces to a boundary theory at finite lattice spacing.
This realizes the anomaly-inflow cancellation mechanism in the bosonized lattice
formulation.

We also gave a prescription for extracting correlation functions of the original
fermion theory from the bosonized theory. This prescription assumes that the
lattice model reaches, in the continuum limit, the theory corresponding to the
target fermionic theory via bosonization.

The formulation developed in this paper is not a direct lattice regularization of
the fermionic chiral gauge theory itself. Rather, it is a lattice formulation of
the bosonic counterpart obtained after introducing spectator fermions. In previous
works based on the Villain formulation~\cite{Villain:1974ir}, the correspondence
between bosons and fermions at the lattice level is controlled by an exact lattice
T-duality~\cite{Berkowitz:2023exact,Thorngren:2026disentangler,
Seifnashri:Exactly-Solvable,Sulejmanpasic:2019with(out)-monopoles,
Gorantla:2021Villain-fracton}. By contrast, it is not straightforward to implement
an analogous duality for the present non-Abelian WZW model on the lattice.
Therefore, whether the lattice theory constructed here flows to the desired continuum fixed
point, and hence to the same universality class as the target fermionic theory,
remains an important question for future study. A detailed analysis of the RG flow is needed for this purpose.

Let us also comment on the scope of the present construction. In this paper, we
have considered only perturbative gauge anomalies and their cancellation. We have
not addressed possible non-perturbative anomalies~\cite{Witten:global-current-algebra,
Gawedzki:2010rn}. The construction was mainly formulated for $SU(N)$ background
gauge fields. Since
\begin{align}
    \pi_1{\left[SU(N)\right]}=\pi_2{\left[SU(N)\right]}=0 ,
\end{align}
the background fields considered in this paper can be treated, in two dimensions,
within the sector continuously connected to the trivial background. In this setting,
the possible obstruction is the perturbative gauge anomaly discussed in the main
text.

This situation changes if the gauge group is no longer connected, or if it
has disconnected components. Examples include $SU(N)/\mathbb{Z}_N$ and finite
groups. In such cases, there can be background gauge fields which are locally
well-defined but cannot be continuously deformed to the trivial background.
Equivalently, there can be global data which are invisible in a purely local
description of the gauge field. Additional topological terms may then be assigned
to such global data, and their gauge variation can give rise to anomalies which
are not captured by the perturbative anomaly polynomial~\cite{Aharony:2013hda,
Abe:2023ncy}.

The present lattice construction relies on choosing continuous interpolations of
lattice fields and on making consistent choices of fractional powers of link
variables%
\footnote{
A possible future direction is to use a non-Abelian Villain-type formulation. Recent
works based on higher-categorical structures propose lattice frameworks for defining Wess--Zumino terms, Chern--Simons terms, and instanton-density operators
on the lattice~\cite{Chen:2024-instanton-density,Zhang-Chen:2024-explicit}.
It would be interesting to reformulate the present construction in such a
framework, where the smoothness conditions and interpolation choices might be
avoided, or at least reorganized.
}
. If the background contains the global data described above, such choices
may not be possible globally, or they may depend on extra discrete choices. This is
a different issue from the perturbative anomaly cancellation studied in this paper.
It should be analyzed separately when one tries to extend the construction to
gauge groups which are not simply connected, gauge groups with disconnected
components, or higher-dimensional chiral gauge theories with nontrivial instanton
sectors.

We finally comment on a possible subtlety of lattice Chern--Simons-type terms.
In attempts to formulate pure Abelian lattice Chern--Simons theory, an additional
zero-mode problem has been discussed in the literature~\cite{Berruto:2000dp,Berruto:2000fb}. In recent formulations of Abelian lattice
Chern--Simons theory and Chern--Simons--Maxwell theory, the Maxwell term also
plays an important role~\cite{DeMarco-CS:2019pqv,Xu-CS:2024hyo,Ikeda:2026lyl}.
It is not clear whether the same issue arises in the non-Abelian construction studied in this paper. Nevertheless, from the viewpoint of Abelian projection
\cite{tHooft-projection:1981bkw,Kronfeld-projection:1987ri}, one may focus on
the Cartan components. A similar zero-mode structure could then appear in the
Abelianized sector. Whether such a zero mode can be lifted by an appropriate
plaquette term should be analyzed separately. In particular, from the
Abelian-projection viewpoint, this plaquette term can be interpreted as a Maxwell
term for the Cartan components. A detailed analysis of this issue is beyond the
scope of the present paper. We therefore leave the zero-mode structure of
non-Abelian lattice Chern--Simons-type terms and its control by plaquette terms
for future work.

Another open problem is the construction of bosonic operators corresponding to
fermion mass terms and Yukawa-type interactions. Clarifying how such interactions
should be represented in the bosonized lattice theory is an important step toward
applications to more general non-Abelian chiral gauge theories.


\acknowledgments
The author would like to thank Hiroshi
Suzuki, Ken Shiozaki, and Yuma Furuta for useful discussions. The author would like
to especially thank Hiroshi Suzuki for a careful reading of the manuscript and providing me insightful comments.
This work was
partially supported by Japan Society for the Promotion of Science (JSPS) Grant-in-Aid for
Scientific Research Grant Numbers JP25KJ1954.

\appendix

\section{Fractional powers of unitary matrices}
\label{app:fractional_power}

In this appendix, we summarize our definition for fractional powers used in the
main text. Let $U\in U(N_f)$. If $U$ has no eigenvalue equal to $-1$, namely when
\begin{align}
    \norm{U-1}<2,
    \label{eq:bound-by-two}
\end{align}
then we define
\begin{align}
    U^s:=\exp\left(s\ln U\right),
    \qquad
    0\leq s\leq 1,
\end{align}
by using the principal logarithm%
\footnote{
The logarithm of a matrix~$U$ is defined as follows. If
\begin{align}
U=\gamma\operatorname{diag}(a_1,\dots,a_N)\gamma^{-1},
\qquad
\gamma\in U(N_f),
\qquad
a_1,\dots,a_N\in\mathbb{C},
\end{align}
then
\begin{align}
\ln U=\gamma\operatorname{diag}(\ln a_1,\dots,\ln a_N)\gamma^{-1}.
\end{align}
}
. In the definition of $\ln U$, the arguments of the eigenvalues are taken in the
principal branch~$(-\pi,\pi)$. This definition satisfies, for $V\in U(N_f)$,
\begin{align}
    U^0=1,\qquad U^1=U,
    \qquad
    (VUV^{-1})^s=VU^sV^{-1},
    \qquad
    (U^{-1})^s=(U^s)^{-1}.
\end{align}

For $U\in SU(N)$, however, we need a definition such that the fractional power
also takes values in $SU(N)$. In the region where $U\in SU(N)$ can be written
uniquely as
\begin{align}
    U=V\operatorname{diag}(e^{\im\theta_1},\dots,e^{\im\theta_N})V^{-1},
    \qquad
    V\in SU(N),
    \qquad
    |\theta_j|<\pi,
    \qquad
    \sum_{j=1}^N \theta_j=0,
\end{align}
we define
\begin{align}
    U^s:=V\operatorname{diag}(e^{\im s\theta_1},\dots,e^{\im s\theta_N})V^{-1}.
\end{align}
A useful sufficient condition for the above definition is
\begin{align}
    \norm{1-U}<2\sin\frac{\pi}{2N},
    \label{eq:sin-bound}
\end{align}
where $\norm{\cdot}$ denotes the operator norm. Indeed, choosing the phases as
$-\pi<\theta_j\leq\pi$, Eq.~\eqref{eq:sin-bound} implies $|\theta_j|<\pi/N$ for
all $j$. Since $\det U=1$, we have $\sum_j\theta_j\in2\pi\mathbb Z$. At the same
time, $|\theta_j|<\pi/N$ gives $|\sum_j\theta_j|<\pi$. Hence
\begin{align}
    \sum_{j=1}^N\theta_j=0 .
\end{align}
Therefore,
\begin{align}
    \det U^s
    =
    \exp\left(\im s\sum_{j=1}^N\theta_j\right)
    =
    1,
\end{align}
and $U^s\in SU(N)$ for any $s\in[0,1]$%
\footnote{
Equivalently, in the region~\eqref{eq:sin-bound}, this definition is the same as
\begin{align}
    U^s=\exp\left(s\ln U\right)
\end{align}
with the principal logarithm.
}.


\section{Sufficient bounds for the smoothness conditions}
\label{app:bounds}

To ensure that all fractional powers used in the main text are uniquely defined,
it is sufficient that their arguments lie in the regions specified by
Eqs.~\eqref{eq:bound-by-two} and~\eqref{eq:sin-bound}. In this appendix, we give
sufficient conditions for this purpose.

In this paper, we impose smoothness conditions on the lattice fields, such as
Eqs.~\eqref{smoothness-condition} and~\eqref{admissibility}. We would like to
choose bounds on the parameters $\varepsilon$ and $\delta$ appearing there so
that $S^L(x;g_1)$ is well-defined. Suppose that we use the
expression~\eqref{eq:S-within-cube}, which can also be used inside a cube, as the
definition of $S^L(x;g_1)$. First, to define an interpolation of
\begin{align}
    \left(g_1(a)^\dagger U_{ab}^Lg_1(b)\right)^x,
\end{align}
we need
\begin{align}
    0<\varepsilon<2
\end{align}
by the bound~\eqref{eq:bound-by-two} for defining fractional powers in $U(N_f)$.
Similarly, to define an interpolation of
\begin{align}
    (U_p)^x,
\end{align}
we need
\begin{align}
    0<\delta<2\sin\frac{\pi}{2N}
\end{align}
by the bound~\eqref{eq:sin-bound} for defining fractional powers in $SU(N)$.
Under these conditions, the fractional powers of the individual link-type and
plaquette-type factors appearing in expressions of the form
\begin{align}
\left(\left[g_1(a)^\dagger U_{ac}^L g_1(c)\right]^{x_\beta}\right)^{-1}
    \,g_1(a)^\dagger
    R_L\left[
        \left(
        U_{ac}U_{cd}U_{db}U_{ba}
        \right)^{x_\beta}
    \right]
    U_{ab}^L g_1(b)
    \left[g_1(b)^\dagger U_{bd}^L g_1(d)\right]^{x_\beta}
\end{align}
are well-defined.

Next, in Eq.~\eqref{eq:S-within-cube}, we need to define interpolations of the
form $\left[ F^L(x_3;g_1) \right]^{x_2}$. This requires fractional powers of
matrices of the type
\begin{align}
\left[\left(\left[g_1(a)^\dagger U_{ac}^L g_1(c)\right]^{x_\beta}\right)^{-1}
    \,g_1(a)^\dagger
    R_L\left[
        \left(
        U_{ac}U_{cd}U_{db}U_{ba}
        \right)^{x_\beta}
    \right]
    U_{ab}^L g_1(b)
    \left[g_1(b)^\dagger U_{bd}^L g_1(d)\right]^{x_\beta}
    \right]^y.
\end{align}
For this to be well-defined, it is sufficient that
\begin{align}
    \norm{\left(\left[g_1(a)^\dagger U_{ac}^L g_1(c)\right]^{x_\beta}\right)^{-1}
    \,g_1(a)^\dagger
    R_L\left[
        \left(
        U_{ac}U_{cd}U_{db}U_{ba}
        \right)^{x_\beta}
    \right]
    U_{ab}^L g_1(b)
    \left[g_1(b)^\dagger U_{bd}^L g_1(d)\right]^{x_\beta}-1}<2 .
\end{align}
Since all building blocks are unitary matrices, the triangle inequality gives
\begin{align}
&\norm{\left(\left[g_1(a)^\dagger U_{ac}^L g_1(c)\right]^{x_\beta}\right)^{-1}
    \,g_1(a)^\dagger
    R_L\left[
        \left(
        U_{ac}U_{cd}U_{db}U_{ba}
        \right)^{x_\beta}
    \right]
    U_{ab}^L g_1(b)
    \left[g_1(b)^\dagger U_{bd}^L g_1(d)\right]^{x_\beta}-1}\notag\\
&\leq\norm{\left[g_1(a)^\dagger U_{ac}^L g_1(c)\right]^{x_\beta}-1}
+\norm{g_1(a)^\dagger U_{ab}^L g_1(b)-1}\notag\\
&\qquad+\norm{\left[g_1(b)^\dagger U_{bd}^L g_1(d)\right]^{x_\beta}-1}
+\norm{R_L\left[
        \left(
        U_{ac}U_{cd}U_{db}U_{ba}
        \right)^{x_\beta}
    \right]-1}\notag\\
&\leq 3\varepsilon
+\norm{R_L\left[
        \left(
        U_{ac}U_{cd}U_{db}U_{ba}
        \right)^{x_\beta}
    \right]-1}.
    \label{remain-rep-bound}
\end{align}
Here we used the following elementary estimate. If the eigenvalues of a unitary
matrix $U$ are $e^{\im\theta_j}$ with $-\pi<\theta_j<\pi$, then the eigenvalues
of $U^x-1$ are $e^{\im x\theta_j}-1$. Hence, for $0\leq x\leq1$,
\begin{align}
    \norm{U^x-1}
    &=
    \max_j \left|e^{\im x\theta_j}-1\right|
    =
    \max_j 2\left|\sin\frac{x\theta_j}{2}\right|
    \notag\\
    &\leq
    \max_j 2\left|\sin\frac{\theta_j}{2}\right|
    =
    \norm{U-1}.
\end{align}

It remains to estimate
\begin{align}
    \norm{R_L\left[
        \left(
        U_{ac}U_{cd}U_{db}U_{ba}
        \right)^{x_\beta}
    \right]-1}.
\end{align}
Let
\begin{align}
    U_p:=U_{ac}U_{cd}U_{db}U_{ba}.
\end{align}
The admissibility condition gives
\begin{align}
    \norm{U_p-1}<\delta<2\sin\frac{\pi}{2N}.
\end{align}
Therefore $U_p$ lies in the region where the principal logarithm is well-defined
as an $SU(N)$ matrix. We write
\begin{align}
    U_p=e^{\im H},
    \qquad
    H=\sum_a H^a t^a,
\end{align}
where $H$ is Hermitian and traceless. If $h_j$ are the eigenvalues of $H$, then
the eigenvalues of $H$ are in the principal branch and
$\norm{H}=\max_j|h_j|<\pi$. Thus,
\begin{align}
    \norm{U_p-1}
    =
    \max_j |e^{\im h_j}-1|
    =
    2\sin\frac{\norm{H}}{2}.
\end{align}
Hence
\begin{align}
    \norm{H}<2\arcsin\left(\frac{\delta}{2}\right).
\end{align}
The fractional power of $U_p$ is then
\begin{align}
    \left(
        U_{ac}U_{cd}U_{db}U_{ba}
        \right)^{x_\beta}
    =
    e^{\im x_\beta H}
    =
    e^{\im x_\beta\sum_a H^a t^a}.
\end{align}
After mapping this to the left representation, we have
\begin{align}
    R_L\left[
        \left(
        U_{ac}U_{cd}U_{db}U_{ba}
        \right)^{x_\beta}\right]
        =
        e^{\im x_\beta\sum_{a}H^at^a_L}.
\end{align}
Therefore,
\begin{align}
    \norm{R_L\left[
        \left(
        U_{ac}U_{cd}U_{db}U_{ba}
        \right)^{x_\beta}
    \right]-1}
    \leq
    \norm{\sum_{a}H^at^a_L}
    \leq
    \sum_{a}|H^a|\norm{t^a_L}.
    \label{eq:rep-plaquette-H-bound}
\end{align}
We use the standard normalization of the generators in the fundamental
representation,
\begin{align}
    \Tr(t^a t^b)=\frac{1}{2}\delta^{ab}.
\end{align}
Then, from $H=\sum_a H^a t^a$, we have
\begin{align}
    H^a=2\Tr(t^aH).
\end{align}
Thus,
\begin{align}
    |H^a|
    =
    2|\Tr(t^aH)|
    \leq
    2N\norm{t^a}\norm{H}.
\end{align}
Combining this inequality with Eq.~\eqref{eq:rep-plaquette-H-bound}, we get
\begin{align}
    \norm{R_L\left[
        \left(
        U_{ac}U_{cd}U_{db}U_{ba}
        \right)^{x_\beta}
    \right]-1}
    \leq
    2N\norm{H}\sum_{a}\norm{t^a}\norm{t^a_L}.
\end{align}
Using the bound on $\norm{H}$, we obtain
\begin{align}
    \norm{R_L\left[
        \left(
        U_{ac}U_{cd}U_{db}U_{ba}
        \right)^{x_\beta}
    \right]-1}
    <
4N\arcsin{\left(\frac{\delta}{2}\right)}
\sum_{a}\norm{t^a}\norm{t^a_L}.
\end{align}
Together with Eq.~\eqref{remain-rep-bound}, this gives
\begin{align}
&\norm{\left(\left[g_1(a)^\dagger U_{ac}^L g_1(c)\right]^{x_\beta}\right)^{-1}
    \,g_1(a)^\dagger
    R_L\left[
        \left(
        U_{ac}U_{cd}U_{db}U_{ba}
        \right)^{x_\beta}
    \right]
    U_{ab}^L g_1(b)
    \left[g_1(b)^\dagger U_{bd}^L g_1(d)\right]^{x_\beta}-1}\notag\\
&<3\varepsilon
+4N\arcsin{\left(\frac{\delta}{2}\right)}
\sum_{a}\norm{t^a}\norm{t^a_L}.
\end{align}
Thus, to define interpolations of the type
\begin{align}
\left[\left(\left[g_1(a)^\dagger U_{ac}^L g_1(c)\right]^{x_\beta}\right)^{-1}
    \,g_1(a)^\dagger
    R_L\left[
        \left(
        U_{ac}U_{cd}U_{db}U_{ba}
        \right)^{x_\beta}
    \right]
    U_{ab}^L g_1(b)
    \left[g_1(b)^\dagger U_{bd}^L g_1(d)\right]^{x_\beta}
    \right]^y,
\end{align}
it is sufficient to impose
\begin{align}
    3\varepsilon
+4N\arcsin{\left(\frac{\delta}{2}\right)}
\sum_{a}\norm{t^a}\norm{t^a_L}<2.
\end{align}

Furthermore, in order to define Eq.~\eqref{eq:S-within-cube}, we also need the
interpolation
\begin{align}
    \left[ J^L(x_2,x_3;g_1) \right]^{x_1}.
\end{align}
The explicit expression of $J^L(x_2,x_3;g_1)$ is
\begin{align}
&J^L(x_2,x_3;g_1)\notag\\
&=
\left[ F^L(x_3;g_1)^{-1} \right]^{x_2}
\left(g_1(3)^\dagger U_{30}^Lg_1(0)\right)^{x_3}
g_1(0)^\dagger Q^L(x_2,x_3)
g_1(1)
\left(g_1(1)^\dagger U_{15}^Lg_1(5)\right)^{x_3}
\left[G^L(x_3;g_1)\right]^{x_2},
\label{eq:J-left-bound-app}
\end{align}
where
\begin{align}
&Q^L(x_2,x_3)\notag\\
&:=
R_L\left[
   \left\{
   \left(U_{03}U_{37}U_{72}U_{20}\right)^{x_3}
  U_{02}
\left(U_{27}U_{74}U_{46}U_{62}\right)^{x_3}
  U_{26}
  U_{61}
  \left(U_{16}U_{64}U_{45}U_{51}\right)^{x_3}
  U_{10}
\left(U_{01}U_{15}U_{53}U_{30}\right)^{x_3}
  \right\}^{x_2}\right.\notag\\
 &\left.\qquad\qquad\qquad\times
 \left(U_{03}U_{35}U_{51}
  U_{10}\right)^{x_3}U_{01}\right].
\label{eq:Q-left-bound-app}
\end{align}
The only new type of fractional power in this expression is
\begin{align}
   \left\{
   \left(U_{03}U_{37}U_{72}U_{20}\right)^{x_3}
  U_{02}
\left(U_{27}U_{74}U_{46}U_{62}\right)^{x_3}
  U_{26}
  U_{61}
\left(U_{16}U_{64}U_{45}U_{51}\right)^{x_3}
  U_{10}
\left(U_{01}U_{15}U_{53}U_{30}\right)^{x_3}
  \right\}^{x_2}.
\end{align}
For this to be well-defined, it is sufficient that
\begin{align}
&\norm{\left(U_{03}U_{37}U_{72}U_{20}\right)^{x_3}
  U_{02}
\left(U_{27}U_{74}U_{46}U_{62}\right)^{x_3}
  U_{26}
  U_{61}
\left(U_{16}U_{64}U_{45}U_{51}\right)^{x_3}
  U_{10}
\left(U_{01}U_{15}U_{53}U_{30}\right)^{x_3}-1}\notag\\
&<2\sin\frac{\pi}{2N}.
\label{eq:Q-core-bound-needed}
\end{align}
This can be bounded in the same way as in the estimate leading to
Eq.~\eqref{remain-rep-bound}. Namely, one repeatedly uses the triangle
inequality, the unitarity of the surrounding factors, and $\norm{P^x-1}\leq\norm{P-1}<\delta$
for each plaquette variable~$P$.
This gives a sufficient bound
\begin{align}
&\norm{\left(U_{03}U_{37}U_{72}U_{20}\right)^{x_3}
  U_{02}
\left(U_{27}U_{74}U_{46}U_{62}\right)^{x_3}
  U_{26}
  U_{61}
\left(U_{16}U_{64}U_{45}U_{51}\right)^{x_3}
  U_{10}
\left(U_{01}U_{15}U_{53}U_{30}\right)^{x_3}-1}\notag\\
&<5\delta.
\end{align}
The numerical coefficient is not important for the following argument; it only
gives a simple sufficient condition. Therefore, Eq.~\eqref{eq:Q-core-bound-needed}
is guaranteed if
\begin{align}
    0<\delta<\frac{2}{5}\sin\frac{\pi}{2N}.
\end{align}

We now estimate $J^L(x_2,x_3;g_1)$. From Eq.~\eqref{eq:J-left-bound-app}, the
triangle inequality gives
\begin{align}
    \norm{J^L(x_2,x_3;g_1)-1}
    &<
    2\left(3\varepsilon
+4N\arcsin{\left(\frac{\delta}{2}\right)}
\sum_{a}\norm{t^a}\norm{t^a_L}\right)
\notag\\
&\qquad
+6\times 4N\arcsin{\left(\frac{\delta}{2}\right)}
\sum_{a}\norm{t^a}\norm{t^a_L}
+3\varepsilon \notag\\
&=9\varepsilon
+32N\arcsin{\left(\frac{\delta}{2}\right)}
\sum_{a}\norm{t^a}\norm{t^a_L}.
\end{align}
Thus, if
\begin{align}
    9\varepsilon
+32N\arcsin{\left(\frac{\delta}{2}\right)}
\sum_{a}\norm{t^a}\norm{t^a_L}<2,
\end{align}
then the interpolation $\left[ J^L(x_2,x_3;g_1) \right]^{x_1}$ is well-defined.

In summary, for the interpolation
\begin{align}
    S^L(x_1,x_2,x_3;g_1)
  =\left(g_1(0)^\dagger U_{03}^Lg_1(3)\right)^{x_3}
  \left[ F^L(x_3;g_1) \right]^{x_2}
  \left[ J^L(x_2,x_3;g_1) \right]^{x_1}
\end{align}
to be well-defined, it is sufficient to choose $\varepsilon$ and $\delta$ in a
common region satisfying
\begin{align}
    &0<\varepsilon<2,\\
    &0<\delta<\frac{2}{5}\sin\frac{\pi}{2N},\\
    &0<3\varepsilon
+4N\arcsin{\left(\frac{\delta}{2}\right)}
\sum_{a}\norm{t^a}\norm{t^a_L}<2,\\
&0<9\varepsilon
+32N\arcsin{\left(\frac{\delta}{2}\right)}
\sum_{a}\norm{t^a}\norm{t^a_L}<2 .
\end{align}
This bound is not optimal. What is needed is that the principal logarithms
required for defining the fractional powers do not cross the branch cut and that
the interpolations are smoothly defined. The bounds above give sufficient
conditions for this purpose.

The same discussion applies to the right sector after replacing $t_L^a$ by
$t_R^a$:
\begin{align}
    &0<3\varepsilon
+4N\arcsin{\left(\frac{\delta}{2}\right)}
\sum_{a}\norm{t^a}\norm{t^a_R}<2,\\
&0<9\varepsilon
+32N\arcsin{\left(\frac{\delta}{2}\right)}
\sum_{a}\norm{t^a}\norm{t^a_R}<2 .
\end{align}
Thus, $\varepsilon$ and $\delta$ should be chosen in the intersection of the
left-sector and right-sector sufficient regions.


\section{An explicit expression of $S^L(x;g_1)$ within a cube}
\label{S^L-within-cube}

In this appendix, we give an explicit formula for $S^L(x;g_1)$ inside a cube~$c$
which satisfies the boundary conditions~\eqref{boundary-SL-123} and
\eqref{boundary-SL-456} on~$\partial c$. The explicit formula presented below is an appropriate modification of L\"uscher's interpolation formula~\cite{Luscher:1981-Topology}. We define $S^L(x;g_1)$ for
$x=(x_1,x_2,x_3)$ inside the cube~$c$ as follows:
\begin{align}
  &F^L(x_3;g_1)\notag\\
  &:=\left(g_1(3)^\dagger U_{30}^Lg_1(0)\right)^{x_3}
  g_1(0)^\dagger R_L\left[(U_{03}U_{37}U_{72}
  U_{20})^{x_3}\right]
  U_{02}^Lg_1(2)\left(g_1(2)^\dagger U_{27}^Lg_1(7)\right)^{x_3},\\
  &G^L(x_3;g_1)\notag\\
  &:=\left(g_1(5)^\dagger U_{51}^Lg_1(1)\right)^{x_3}
  g_1(1)^\dagger R_L\left[\left(U_{15}U_{54}U_{46}
  U_{61}\right)^{x_3}\right]
  U_{16}^Lg_1(6)\left(g_1(6)^\dagger U_{64}^Lg_1(4)\right)^{x_3},\\
&Q^L(x_2,x_3)\notag\\
&:=
R_L\left[
   \left\{
   \left(U_{03}U_{37}U_{72}U_{20}\right)^{x_3}
  U_{02}
\left(U_{27}U_{74}U_{46}U_{62}\right)^{x_3}
  U_{26}
  U_{61}
  \left(U_{16}U_{64}U_{45}U_{51}\right)^{x_3}
  U_{10}
\left(U_{01}U_{15}U_{53}U_{30}\right)^{x_3}
  \right\}^{x_2}\right.\notag\\
 &\left.\qquad\qquad\qquad\times 
 \left(U_{03}U_{35}U_{51}
  U_{10}\right)^{x_3}U_{01}\right],\\
&J^L(x_2,x_3;g_1)\notag\\
  &:=\left[ F^L(x_3;g_1)^{-1} \right]^{x_2}
  \left(g_1(3)^\dagger U_{30}^Lg_1(0)\right)^{x_3}
  g_1(0)^\dagger Q^L(x_2,x_3)
  g_1(1)\left(g_1(1)^\dagger U_{15}^Lg_1(5)\right)^{x_3}
  \left[G^L(x_3;g_1)\right]^{x_2},\\
  &S^L(x_1,x_2,x_3;g_1)
  :=\left(g_1(0)^\dagger U_{03}^Lg_1(3)\right)^{x_3}
  \left[ F^L(x_3;g_1) \right]^{x_2}
  \left[ J^L(x_2,x_3;g_1) \right]^{x_1}.
  \label{eq:S-within-cube}
\end{align}
We now check that this expression satisfies the boundary conditions
\eqref{boundary-SL-123} and~\eqref{boundary-SL-456} on~$\partial c$.

For $x\in(0273)$, namely $x_1=0$, we obtain
\begin{align}
    &S^L(0,x_2,x_3;g_1)\notag\\
  &=\left(g_1(0)^\dagger U_{03}^Lg_1(3)\right)^{x_3}
  \left[ \left(g_1(3)^\dagger U_{30}^Lg_1(0)\right)^{x_3}
  g_1(0)^\dagger R_L\left[(U_{03}U_{37}U_{72}
  U_{20})^{x_3}\right]
  U_{02}^Lg_1(2)\left(g_1(2)^\dagger U_{27}^Lg_1(7)\right)^{x_3}\right]^{x_2}.
\end{align}
For $x\in(1645)$, namely $x_1=1$, we obtain
\begin{align}
    &S^L(1,x_2,x_3;g_1)\notag\\
    &=\left(g_1(0)^\dagger U_{03}^Lg_1(3)\right)^{x_3}
  \left[ F^L(x_3;g_1) \right]^{x_2}
  J^L(x_2,x_3;g_1)
\notag\\
&=g_1(0)^\dagger Q^L(x_2,x_3)
  g_1(1)\notag\\
  &\times\left(g_1(1)^\dagger U_{15}^Lg_1(5)\right)^{x_3}
  \left[\left(g_1(5)^\dagger U_{51}^Lg_1(1)\right)^{x_3}
  g_1(1)^\dagger R_L\left[\left(U_{15}U_{54}U_{46}
  U_{61}\right)^{x_3}\right]
  U_{16}^Lg_1(6)\left(g_1(6)^\dagger U_{64}^Lg_1(4)\right)^{x_3}\right]^{x_2}.
\end{align}
For $x\in(0153)$, namely $x_2=0$, we obtain
\begin{align}
    &S^L(x_1,0,x_3;g_1)\notag\\
  &=\left(g_1(0)^\dagger U_{03}^Lg_1(3)\right)^{x_3}
  \left[ \left(g_1(3)^\dagger U_{30}^Lg_1(0)\right)^{x_3}
  g_1(0)^\dagger Q^L(0,x_3)
  g_1(1)\left(g_1(1)^\dagger U_{15}^Lg_1(5)\right)^{x_3} \right]^{x_1}
  \notag\\
&=\left(g_1(0)^\dagger U_{03}^Lg_1(3)\right)^{x_3}
  \left[ \left(g_1(3)^\dagger U_{30}^Lg_1(0)\right)^{x_3}
  g_1(0)^\dagger 
R_L\left[\left(U_{03}U_{35}U_{51}
  U_{10}\right)^{x_3}U_{01}\right]
  g_1(1)\left(g_1(1)^\dagger U_{15}^Lg_1(5)\right)^{x_3} \right]^{x_1}.
\end{align}
For $x\in(2647)$, namely $x_2=1$, we obtain
\begin{align}
    &S^L(x_1,1,x_3;g_1)\notag\\
  &=\left(g_1(0)^\dagger U_{03}^Lg_1(3)\right)^{x_3}
  \left[ F^L(x_3;g_1) \right]
  \left[ J^L(1,x_3;g_1) \right]^{x_1}\notag\\
  &=g_1(0)^\dagger R_L\left[(U_{03}U_{37}U_{72}
  U_{20})^{x_3}\right]U_{02}^Lg_1(2)\notag\\
  &\times
  \left(g_1(2)^\dagger U_{27}^Lg_1(7)\right)^{x_3}\left[ \left(g_1(7)^\dagger U_{72}^Lg_1(2)\right)^{x_3}g_1(2)^\dagger
  R_L\left[
\left(U_{27}U_{74}U_{46}U_{62}\right)^{x_3}
  U_{26}\right]
  g_1(6)\left(g_1(6)^\dagger U_{64}^Lg_1(4)\right)^{x_3} \right]^{x_1}.
\end{align}
For $x\in(0162)$, namely $x_3=0$, we obtain
\begin{align}
    &S^L(x_1,x_2,0;g_1)\notag\\
  &=
  \left[ F^L(0;g_1) \right]^{x_2}
  \left[ J^L(x_2,0;g_1) \right]^{x_1}\notag\\
  &=
  \left[ g_1(0)^\dagger U_{02}^Lg_1(2) \right]^{x_2}
  \left[ 
  \left[ g_1(2)^\dagger U_{20}^Lg_1(0) \right]^{x_2}
  g_1(0)^\dagger 
  R_L\left[
   \left\{   
U_{02}U_{26}U_{61}U_{10}
\right\}^{x_2}U_{01}\right]
  g_1(1)
  \left[g_1(1)^\dagger U_{16}^Lg_1(6)\right]^{x_2}
  \right]^{x_1}.
\end{align}
For $x\in(3547)$, namely $x_3=1$, we obtain
\begin{align}
    &S^L(x_1,x_2,1;g_1)\notag\\
  &=g_1(0)^\dagger U_{03}^Lg_1(3)
  \left[ F^L(1;g_1) \right]^{x_2}
  \left[ J^L(x_2,1;g_1) \right]^{x_1}\notag\\
  &=g_1(0)^\dagger U_{03}^Lg_1(3)
  \left[ g_1(3)^\dagger U_{37}^Lg_1(7) \right]^{x_2}
  \left[ J^L(x_2,1;g_1) \right]^{x_1}\notag\\
  &=g_1(0)^\dagger U_{03}^Lg_1(3)\notag\\
  &\times
  \left[ g_1(3)^\dagger U_{37}^Lg_1(7) \right]^{x_2}
  \left[ \left[ g_1(7)^\dagger U_{73}^Lg_1(3) \right]^{x_2}g_1(3)^\dagger
  R_L\left[
   \left\{
   U_{37}U_{74}
  U_{45}U_{53}
  \right\}^{x_2} 
 U_{35}
  \right]
  g_1(5)
  \left[g_1(5)^\dagger U_{54}^Lg_1(4)\right]^{x_2} \right]^{x_1}.
\end{align}


\section{Details for the interpolation~$\mathrm{Int}_x[\cdot]$}
\label{sec:def-Int}

In this appendix, we define the interpolation $\mathrm{Int}_x[\cdot]$.
For plaquette-type closed products, we use fractional powers which are uniquely
defined by the admissibility condition~\eqref{admissibility}. On the other hand,
when a single link variable itself is connected continuously to the identity, and
when a gauge-covariant and unique definition of a fractional power is not needed,
we choose one interpolation for each fixed link configuration.

The parameter spaces of the families to which $\mathrm{Int}$ is applied below
are at most two-dimensional. Moreover, for $N\geq2$,
\begin{align}
    \pi_i(SU(N))=0,
    \qquad
    i=0,1,2.
\end{align}
Therefore, the interpolations needed below exist. We denote such
an interpolation by
\begin{align}
    \mathrm{Int}_{x}[A].
\end{align}
This notation does not represent a global map defined on the whole group
$SU(N)$. Rather, it represents one interpolation chosen for each
$SU(N)$-valued family appearing below, after fixing the link configuration.

We first fix interpolations in the $x_3$ direction for the oriented link
variables
\begin{align}
    U_{30},\quad U_{27},\quad U_{51},\quad
    U_{64},\quad U_{03},\quad U_{15}.
\end{align}
We denote them by
\begin{align}
    \mathrm{Int}_{x_3}[U_{30}],\quad
    \mathrm{Int}_{x_3}[U_{27}],\quad
    \mathrm{Int}_{x_3}[U_{51}],\quad
    \mathrm{Int}_{x_3}[U_{64}],\quad
    \mathrm{Int}_{x_3}[U_{03}],\quad
    \mathrm{Int}_{x_3}[U_{15}].
\end{align}
Each interpolation satisfies
\begin{align}
    \mathrm{Int}_{0}[U_{ab}]=1,
    \qquad
    \mathrm{Int}_{1}[U_{ab}]=U_{ab}.
\end{align}
For oppositely oriented links, we do not choose interpolations independently.
Instead, we define
\begin{align}
    \mathrm{Int}_{x_3}[U_{ba}]
    :=
    \left(\mathrm{Int}_{x_3}[U_{ab}]\right)^{-1}.
\end{align}

Using this choice, we define
\begin{align}
  F_I(x_3;1)
  &:=
  \mathrm{Int}_{x_3}\left[U_{30}\right]\,
  \left(U_{03}U_{37}U_{72}U_{20}\right)^{x_3}
  U_{02}\,
  \mathrm{Int}_{x_3}\left[ U_{27}\right],
  \\
  G_I(x_3;1)
  &:=
  \mathrm{Int}_{x_3}\left[U_{51}\right]\,
  \left(U_{15}U_{54}U_{46}U_{61}\right)^{x_3}
  U_{16}\,
  \mathrm{Int}_{x_3}\left[U_{64}\right].
\end{align}

Next, for the families $F_I(x_3;1)$ and $G_I(x_3;1)$ parametrized by $x_3$, we
choose interpolations in the $x_2$ direction. For the inverse, we do
not choose an interpolation independently. Instead, we define
\begin{align}
  \mathrm{Int}_{x_2}\left[ F_I(x_3;1)^{-1} \right]
  :=
  \left(
  \mathrm{Int}_{x_2}\left[ F_I(x_3;1) \right]
  \right)^{-1}.
\end{align}
With this convention, we define
\begin{align}
J_I(x_2,x_3;1)
  &:=
  \mathrm{Int}_{x_2}\left[ F_I(x_3;1)^{-1} \right]\,
  \mathrm{Int}_{x_3}\left[U_{30}\right]\,
  Q(x_2,x_3)\,
  \mathrm{Int}_{x_3}\left[U_{15}\right]\,
  \mathrm{Int}_{x_2}\left[G_I(x_3;1)\right].
\end{align}

Finally, for the $SU(N)$-valued family
\begin{align}
    J_I(x_2,x_3;1)
\end{align}
parametrized by $(x_2,x_3)$, we choose an interpolation in the $x_1$
direction and define
\begin{align}
  S_I(x_1,x_2,x_3;1)
  &:=
  \mathrm{Int}_{x_3}\left[ U_{03}\right]\,
  \mathrm{Int}_{x_2}\left[ F_I(x_3;1) \right]\,
  \mathrm{Int}_{x_1}\left[ J_I(x_2,x_3;1) \right].
\end{align}

The interpolations in the left and right representations are defined by
\begin{align}
  S_I^{L/R}(x_1,x_2,x_3;1)
  &:=
  R_{L/R}\left[S_I(x_1,x_2,x_3;1)\right].
\end{align}
Since $R_{L/R}$ are group homomorphisms, a smooth $SU(N)$-valued interpolation
$S_I$ gives smooth interpolations $S_I^{L/R}$ on the corresponding representation
spaces.

Using the interpolation $\mathrm{Int}_x[\cdot]$ and the fractional
powers defined above, the explicit expression for $S_I(x;1)$ inside a cube is
given by
\begin{align}
  &F_I(x_3;1)
  =\mathrm{Int}_{x_3}\left[U_{30}\right]
\left(U_{03}U_{37}U_{72}U_{20}\right)^{x_3}
U_{02}\,\mathrm{Int}_{x_3}\left[ U_{27}\right],\\
  &G_I(x_3;1)
  =\mathrm{Int}_{x_3}\left[U_{51}\right]
\left(U_{15}U_{54}U_{46}
U_{61}\right)^{x_3}U_{16}\,\mathrm{Int}_{x_3}\left[U_{64}\right],\\
&Q(x_2,x_3)\notag\\
&=
\left\{
   \left(U_{03}U_{37}U_{72}U_{20}\right)^{x_3}
  U_{02}
\left(U_{27}U_{74}U_{46}U_{62}\right)^{x_3}
  U_{26}
  U_{61}
  \left(U_{16}U_{64}U_{45}U_{51}\right)^{x_3}
  U_{10}
\left(U_{01}U_{15}U_{53}U_{30}\right)^{x_3}
  \right\}^{x_2}\notag\\
 &\qquad\qquad\qquad\times 
 \left(U_{03}U_{35}U_{51}
  U_{10}\right)^{x_3}U_{01},\\
&J_I(x_2,x_3;1)
  =\mathrm{Int}_{x_2}\left[ F_I(x_3;1)^{-1} \right]\mathrm{Int}_{x_3}\left[U_{30}\right]
 Q(x_2,x_3)
  \mathrm{Int}_{x_3}\left[U_{15}\right]
  \mathrm{Int}_{x_2}\left[G_I(x_3;1)\right],\\
  &S_I(x_1,x_2,x_3;1)
  =\mathrm{Int}_{x_3}\left[ U_{03}\right]
  \mathrm{Int}_{x_2}\left[ F_I(x_3;1) \right]
  \mathrm{Int}_{x_1}\left[ J_I(x_2,x_3;1) \right],\\
  &S_I^{L/R}(x_1,x_2,x_3;1):=
  R_{L/R}\left[S_I(x_1,x_2,x_3;1)\right].
\end{align}
The field $S_I^{L/R}(x_1,x_2,x_3;1)$ satisfies, on~$\partial c$, the boundary
conditions
\begin{align}
P_{i}^L(x_\alpha,x_\beta;1,I)
&:=
R_L\left[
    \mathrm{Int}_{x_\beta}[U_{ac}]
\right.
\notag\\
&\qquad
\times
\left.
\mathrm{Int}_{x_\alpha}
\left[
    \left(
    \mathrm{Int}_{x_\beta}[U_{ac}]
    \right)^{-1}\,
\left(
    U_{ac}U_{cd}U_{db}U_{ba}
\right)^{x_\beta}
U_{ab}\,
\mathrm{Int}_{x_\beta}[U_{bd}]
\right]
\right],
\\
S^L_I(x;1)&:=P_i^L(x_\alpha,x_\beta;1,I)
    \qquad \text{for } i=1,2,3,
\\
S^L_I(x;1)&:=R_i^L(x;1)\,P_i^L(x_\alpha,x_\beta;1,I)
    \qquad \text{for } i=4,5,6.
\end{align}
This can be verified by a calculation analogous to that in
Appendix~\ref{S^L-within-cube}.


\section{Continuity of the cube interpolation~$g_1^c(x)$}
\label{app:continuity-g1c}

In this appendix, we show that
\begin{align}
    g_1^c(x)
    :=
    \left(S_I^L(x;1)\right)^{-1}g_1(0)S^L(x;g_1)
    \label{eq:g1c-app}
\end{align}
introduced in the main text is continuously glued on the faces, links, and sites
of neighboring cubes. The same argument applies to the right sector after the
replacement $g_1\to g_2^\dagger$ and $L\to R$.

A cube has six faces. On each face, the definitions of $S_I^L(x;1)$ and
$S^L(x;g_1)$ can be written in the following form. For faces $i=1,2,3$, we have
\begin{align}
    S_I^L(x;1)
    &=
    P_i^L(x_\alpha,x_\beta;1,I),
&
    S^L(x;g_1)
    &=
    P_i^L(x_\alpha,x_\beta;g_1).
\end{align}
Therefore,
\begin{align}
    g_1^c(x)
    &=
    \left(P_i^L(x_\alpha,x_\beta;1,I)\right)^{-1}
    g_1(0)
    P_i^L(x_\alpha,x_\beta;g_1).
\end{align}
On the other hand, for faces $i=4,5,6$, we have
\begin{align}
    S_I^L(x;1)
    &=
    R_i^L(x;1)\,P_i^L(x_\alpha,x_\beta;1,I),
&
    S^L(x;g_1)
    &=
    R_i^L(x;g_1)\,P_i^L(x_\alpha,x_\beta;g_1).
\end{align}
Hence,
\begin{align}
    g_1^c(x)
    &=
    \left(P_i^L(x_\alpha,x_\beta;1,I)\right)^{-1}
    g_1(1)
    P_i^L(x_\alpha,x_\beta;g_1),
    \qquad i=4,
\\
    g_1^c(x)
    &=
    \left(P_i^L(x_\alpha,x_\beta;1,I)\right)^{-1}
    g_1(2)
    P_i^L(x_\alpha,x_\beta;g_1),
    \qquad i=5,
\\
    g_1^c(x)
    &=
    \left(P_i^L(x_\alpha,x_\beta;1,I)\right)^{-1}
    g_1(3)
    P_i^L(x_\alpha,x_\beta;g_1),
    \qquad i=6.
\end{align}
Thus, for $x\in(abcd)$, we can write in general
\begin{align}
    g_1^c(x)
    =
    \left(P_i^L(x_\alpha,x_\beta;1,I)\right)^{-1}
    g_1(a)
    P_i^L(x_\alpha,x_\beta;g_1).
    \label{eq:g1c-face-general}
\end{align}
By construction, $P_i^L(x_\alpha,x_\beta;1,I)$ and
$P_i^L(x_\alpha,x_\beta;g_1)$ depend only on the degrees of freedom defined on the
face~$(abcd)$. Hence the restriction of $g_1^c(x)$ to a common face is determined
only by the data on that face. Therefore, two neighboring cubes give the same
value on their common face.

We next check the compatibility on links. We first set $x_\alpha=0$. Then
\begin{align}
    P_i^L(0,x_\beta;1,I)
    &=
    R_L\left[\mathrm{Int}_{x_\beta}[U_{ac}]\right],
\\
    P_i^L(0,x_\beta;g_1)
    &=
    \left[g_1(a)^\dagger U_{ac}^L g_1(c)\right]^{x_\beta}.
\end{align}
Therefore, Eq.~\eqref{eq:g1c-face-general} gives
\begin{align}
    g_1^c(x)
    &=
    \left(R_L\left[\mathrm{Int}_{x_\beta}[U_{ac}]\right]\right)^{-1}
    g_1(a)
    \left[g_1(a)^\dagger U_{ac}^L g_1(c)\right]^{x_\beta}.
    \label{eq:g1c-edge-ac}
\end{align}
This expression depends only on the site variables at the endpoints of the link~$(ac)$ and the link variable on that link.

Next, we set $x_\alpha=1$. In this case,
\begin{align}
    P_i^L(1,x_\beta;1,I)
    &=
    R_L\left[
    \left(
        U_{ac}U_{cd}U_{db} U_{ba}
    \right)^{x_\beta}
    U_{ab}
    \mathrm{Int}_{x_\beta}[U_{bd}]
    \right],
\\
    P_i^L(1,x_\beta;g_1)
    &=
    g_1(a)^\dagger
    R_L\left[
    \left(
        U_{ac} U_{cd} U_{db} U_{ba}
    \right)^{x_\beta}
    U_{ab}
    \right]
    g_1(b)
    \left[g_1(b)^\dagger U_{bd}^L g_1(d)\right]^{x_\beta}.
\end{align}
Substituting these expressions into Eq.~\eqref{eq:g1c-face-general}, we obtain
\begin{align}
    g_1^c(x)
    &=
    \left(R_L\left[
    \mathrm{Int}_{x_\beta}[U_{bd}]\right]\right)^{-1}
    g_1(b)
    \left[g_1(b)^\dagger U_{bd}^L g_1(d)\right]^{x_\beta}.
    \label{eq:g1c-edge-bd}
\end{align}
Thus, the value on $x_\alpha=1$ is determined only by the degrees of freedom on
the link~$(bd)$.

Similarly, let $x_\beta=0$. Then
\begin{align}
    P_i^L(x_\alpha,0;1,I)
    &=
    R_L\left[\mathrm{Int}_{x_\alpha}[U_{ab}]\right],
\\
    P_i^L(x_\alpha,0;g_1)
    &=
    \left[
        g_1(a)^\dagger U_{ab}^L g_1(b)
    \right]^{x_\alpha}.
\end{align}
Therefore,
\begin{align}
    g_1^c(x)
    &=
    \left(R_L\left[\mathrm{Int}_{x_\alpha}[U_{ab}]\right]\right)^{-1}
    g_1(a)
    \left[
        g_1(a)^\dagger U_{ab}^L g_1(b)
    \right]^{x_\alpha}.
    \label{eq:g1c-edge-ab}
\end{align}
This expression is written only in terms of the degrees of freedom on the
link~$(ab)$.

Finally, let $x_\beta=1$. Then
\begin{align}
    P_i^L(x_\alpha,1;1,I)
    &=
    R_L\left[U_{ac}
    \mathrm{Int}_{x_\alpha}[U_{cd}]\right],
\\
    P_i^L(x_\alpha,1;g_1)
    &=
    g_1(a)^\dagger U_{ac}^L g_1(c)\,
    \left[
        g_1(c)^\dagger U_{cd}^L g_1(d)
    \right]^{x_\alpha}.
\end{align}
Substituting these expressions into Eq.~\eqref{eq:g1c-face-general}, we obtain
\begin{align}
    g_1^c(x)
    &=
    \left(R_L\left[
    \mathrm{Int}_{x_\alpha}[U_{cd}]\right]\right)^{-1}
    g_1(c)
    \left[
        g_1(c)^\dagger U_{cd}^L g_1(d)
    \right]^{x_\alpha}.
    \label{eq:g1c-edge-cd}
\end{align}
Hence, the value on $x_\beta=1$ is determined only by the degrees of freedom on
the link~$(cd)$.

We have shown that the restrictions of $g_1^c(x)$ to the links have the common
form
\begin{align}
    g_1^c(x)|_{x\in(pq)}
    =
    \left(R_L\left[\mathrm{Int}_{x}[U_{pq}]\right]\right)^{-1}
    g_1(p)
    \left[
        g_1(p)^\dagger U_{pq}^L g_1(q)
    \right]^{x},
    \qquad 0\leq x\leq 1,
    \label{eq:g1c-edge-universal}
\end{align}
on any oriented link~$(pq)$. This expression is defined only in terms of the data
on the link~$(pq)$. Therefore, it gives the same value independently of which face
sharing the link~$(pq)$ is used.

At the endpoints, Eq.~\eqref{eq:g1c-edge-universal} immediately gives
\begin{align}
    g_1^c(p)
    &=
    g_1(p),
&
    g_1^c(q)
    &=
    \left(U_{pq}^L\right)^{-1}
    g_1(p)
    g_1(p)^\dagger U_{pq}^L g_1(q)
    =
    g_1(q).
\end{align}
Therefore, at the four sites~$a$, $b$, $c$, $d$, we have
\begin{align}
    g_1^c(a)=g_1(a),
    \qquad
    g_1^c(b)=g_1(b),
    \qquad
    g_1^c(c)=g_1(c),
    \qquad
    g_1^c(d)=g_1(d).
\end{align}
Thus, the compatibility at sites is also ensured.

We have therefore shown that the local expressions for $g_1^c(x)$ are
compatible on common faces, common links, and common sites of neighboring
cubes. Consequently, $g_1^c(x)$ defines a well-defined continuous map on the
whole lattice, which is piecewise smooth whenever the interpolation functions
used in the construction are smooth.

\bibliographystyle{JHEP}
\bibliography{references}

\end{document}